\documentclass{aa} 

\makeatletter
\@ifclasswith{aa}{referee}{%
  \newcommand{\docclassaareferee}{true}
}{
  \newcommand{\docclassaareferee}{false}
}
\makeatother

\usepackage{ifthen}
\newlength{\figwidthl}
\ifthenelse{\equal{\docclassaareferee}{true}}{
  \setlength{\figwidthl}{1\columnwidth}
}{
  \setlength{\figwidthl}{1.47\columnwidth}
}

\newlength{\figwidths}
\ifthenelse{\equal{\docclassaareferee}{true}}{
  \setlength{\figwidths}{0.5\columnwidth}
}{
  \setlength{\figwidths}{0.625\columnwidth}
}

\usepackage{graphicx}
\usepackage{txfonts}
\usepackage{siunitx}
\usepackage{gensymb}
\usepackage{subfig}
\usepackage{CJKutf8}
\newcommand{\CNnames}[1]{{\begin{CJK}{UTF8}{gbsn}~(#1)~\end{CJK}}}
\usepackage{multirow}
\usepackage{hyperref}
\hypersetup{
    colorlinks=true,
    linkcolor=blue,
    citecolor=blue,
    urlcolor=blue
    }

\newcommand{\Teff}{{T_\mathrm{eff}}}

\def\angstrom{\text {Å}}

\begin{document} 

   \title{Exploring Be phenomena in OBA stars: a Mid-infrared search\thanks{Tables \ref{tab:bep-catalogue} and \ref{tab:lamost-catalogue} are only available in electronic form at the CDS via anonymous ftp to \url{cdsarc.u-strasbg.fr} (\href{ftp://130.79.128.5/}{130.79.128.5}), or via \url{http://cdsweb.u-strasbg.fr/cgi-bin/qcat?J/A+A/}. The binned light curve for BeT stars and the figures similar to Figure~\ref{fig:lc_lasmost_example}, \ref{fig:lc_golden_example} and \ref{fig:lc_extended_example} for all the BeT stars are only available in electronic form at \url{https://github.com/MingjieJian/wisebet}.}}
   
   \subtitle{}

   \author{Mingjie Jian \CNnames{简明杰}
   \inst{1}\and
           Noriyuki Matsunaga \CNnames{松永典之}
	   \inst{2}\and
           Biwei Jiang \CNnames{姜碧沩}
	   \inst{3}\and
           Haibo Yuan \CNnames{苑海波}
	   $^3$
           \and 
           Ruoyi Zhang \CNnames{张若羿}
	   $^3$
          }
    \titlerunning{WISE Be phenomena}
   \authorrunning{M. Jian et al.}
          
   \institute{Department of Astronomy, Stockholm University, AlbaNova University Center, Roslagstullsbacken 21, 114 21 Stockholm, Sweden\\
              \email{mingjie.jian@astro.su.se}
         \and
             Department of Astronomy, School of Science, The University of Tokyo, 7-3-1 Hongo, Bunkyo-ku, Tokyo 113-0033, Japan
             \and
             Department of Astronomy, Beijing Normal University, Beijing 100875, People's Republic of China\\
             }

   \date{Received September 08, 2023; accepted November 10, 2023}

 
  \abstract
   {As early-type stars with a rotation speed close to their critical velocity, Be stars experience an event called the Be phenomenon.
    The material in their equator is ejected into outside space during the Be phenomenon and forms a circumstellar disk.
    The mechanism triggering these events remains poorly understood, and observations of these events are limited because the duration of these events ranges from months to years.
    Long-term epoch photometry in the infrared bands is expected to be ideal for detecting Be phenomena because the brightness variation is larger than that in the optical, and the effect of interstellar extinction is weaker as well. 
    We conducted a systematic search for Be phenomena among Milky Way OBA stars in the mid-infrared. 
    We examined the brightness and colour variations of known classical Be stars using the WISE $W1$ and $W2$ photometry bands to quantify their characteristics. 
    Subsequently, we established a set of criteria to identify similar photometric variations in a large sample of OBA stars. 
    We found 916 OBA stars that show Be phenomena in the past 13 years, 736 of which are newly discovered. 
    The peak-to-peak variations in magnitude and colour were found to be correlated, indicating that a decretion disk is common. 
    The increase in colour was observed to be strongly correlated with the emission of the H$\alpha$ line, providing further evidence of the association with circumstellar disks. 
    The brightness variation of a star with Be phenomena can be up to \SI{1.5}{mag}, and the colour variations can be up to \SI{0.4}{mag}. 
    The median durations for the disk build-up and decay phases are 474 and 524 days, respectively (durations shorter than 180\,days are not sampled). 
    The search for Be phenomena in the WISE bands greatly enlarges the number of stars showing disk variation, and it enables multi-band photometry analysis of these events with the help of current and future optical photometry surveys.}


   \maketitle
%

\section{Introduction}
\label{sec:intro}

Classical Be stars, which are rapidly rotating B-type stars, exhibit or have exhibited emission in one or more of their Balmer lines \citep{Collins1987}. 
These emission lines serve as signatures of the circumstellar envelope and typically take the form of a disk. 
This disk is thought to have formed during the Be phenomenon \citep{Smith1999, Neiner2013}, where material from the stellar equator is lifted up and forms the so-called decretion disk \citep{Okazaki2001}. 
The decretion disk can be effectively described using the viscous decretion disk (VDD) theory, assuming a viscous Keplerian disk model (see the review of \citealt{Rivinius2013}).
\citet{Haubois2012} demonstrated that the brightness of the star changes throughout the build-up and dissipation phases of the decretion disk. 
When viewed face-on, both the star and the disk contribute to the observed brightness. 
Consequently, the build-up of the decretion disk increases the brightness of the star in both the optical and infrared.
When viewed edge-on, a portion of the stellar light is obstructed by the disk.
The optical brightness may remain stable or even decrease when a decretion disk forms, while the infrared brightness increases because the main radiation of the disk is in the infrared.

The observation of stars with Be phenomena, that is, those that tansit from the B to the Be phase or vice versa (BeT stars) is limited.
Photometry data with a time span of months to years have only become available recently. They correspond to the timescale of the Be phenomenon.
\citet{Mennickent2002} identified 293 Be stars in the Small Magellanic Cloud with outbursts using the Optical Gravitational Lensing Experiment (OGLE; \citealt{Udalski1997}) photometric data, \citet{Labadie-Bartz2017} found 219 Be stars using the Kilodegree Extremely Little Telescope (KELT) data, and \citet{Bernhard2018} identified 102 Be stars using All-Sky Automated Survey-3 data (ASAS-3; \citealt{Pojmanski2002}).
The sample sizes in these studies are still limited by the sky coverage and/or the survey time span.
Moreover, these studies relied on single-band optical photometry, making it challenging to identify BeT stars in regions with significant extinction.
Recently, alternative methods for detecting BeT stars have been explored. 
For example, \citet{Granada2018} used the variability of the Wide-field Infrared Survey Explorer (WISE; \citealt{Wright2010}) photometry and concluded that about half of the Be stars in a few open clusters are active Be stars, that is, have a decretion disk, and a similar approach was applied to Gaia data \citep{Granada2023}.

We aim to reveal the potential of infrared photometry in detecting BeT stars in this study.
The brightness variations during disk build-up or dissipation are expected to be more prominent in the infrared than in the optical, as demonstrated by \citet{Haubois2012}.
Additionally, some early-type stars are located in regions with significant extinction, making it difficult to detect Be phenomena using optical wavelengths alone.
Recently, three Be stars with outbursts were reported in \citet{Froebrich2023}.
These Be stars are brighter in the WISE bands and exhibit a redder $(W1-W2)$ colour during their outburst phase. 
Hence, it may be possible to detect Be phenomena using WISE epoch photometry alone. 
\citet{Mei2023} derived a variability sample for a set of pre-main-sequence (PMS) stars and classical Be stars (CBe).
Of the 8470 PMS stars and 693 CBe stars reported in \citet{Vioque2020}, a subset of 4756 PMS stars and 459 CBe stars exhibited variability in the WISE bands.
These studies validated the possibility of using WISE data to detect Be star variations.
However, the detection criteria employed by \citet{Mei2023} only considered variations in magnitude. 
By excluding the assessment of colour variations, this sample may also include other types of variations in addition to Be phenomena. 
Moreover, the sample in \citet{Mei2023} is largely confined to the northern part of the Galactic plane.
Thus, searching for BeT stars using the WISE magnitude as well as colour across a broader sky coverage is necessary to achieve a better understanding of the statistical characteristics of the Be phenomena and to provide constraints on the triggering mechanisms.

In this study, we present the search for the Be phenomenon in Galactic early-type stars using WISE epoch photometry. 
Section~\ref{sec:data} introduces our data, including the input target list and the WISE photometry. The method for detecting the Be phenomenon, along with manual classification, is described in Section~\ref{sec:wise-lc}. We then use the statistical properties of the BeT stars to establish criteria for detecting them without the need for manual classification, which is subsequently applied to a larger sample (Section~\ref{sec:wise-lc}). 
The detected WISE BeT stars are presented in Section~\ref{sec:result}, followed by a discussion in Section~\ref{sec:discussion}.
Finally, Section~\ref{sec:conclusion} concludes the paper.

\section{Data and catalogue}
\label{sec:data}

\subsection{WISE epoch photometry}
\label{sec:wise-intro}

The WISE satellite is designed to perform all-sky imaging and photometry in the mid-infrared, that is, \num{3.4}, \num{4.6}, \num{12}, and \SI{22}{\micro m} (bands $W1$, $W2$, $W3$, and $W4$).
The satellite surveyed the whole sky from January 7 to August 6, 2010 (the four-band cryo period), before the solid hydrogen in the outer cryogen tank was exhausted.
Later, the detector was warmed from \SI{12}{K} to \SI{45}{K}, and the $W4$ band data were unusable due to the strong thermal emission. 
This is called the three-band cryo period, and it lasted until September 29, 2010, surveying 30\% of the sky.
As the solid hydrogen in the inner cryogen tank was then also exhausted, $W3$ band data taken later on also became unusable, leaving only $W1$ and $W2$ band data.
The telescope continued in operation for another four months until February 1, 2011 (post-cryo period), completing the survey of the remaining 70\% of the sky.
After nearly three years of silence, WISE was reactivated, continued its all-sky survey for near-Earth objects, and was renamed Near-Earth Object Wide-field Infrared Survey Explorer Reactivation Mission (NEOWISE) \citealt{Mainzer2011}).
The mission is ongoing, and it surveyed the whole sky nearly 18 times so far, with $\approx$\SI{180}{days} of separation between the visits and 12 or more independent exposures per visit.
This survey strategy provides time-series photometry data covering a time span of more than ten years in $W1$ and $W2$, making it an ideal dataset for detecting short-term (shorter than 15 days) or long-term (e.g. longer than six months) brightness variation for stars.

Several previous studies have demonstrated the power of WISE epoch photometry. 
For example, \citet{Chen2018} discovered more than 30,000 variable stars, including 1312 Cepheids, using the phase-folded WISE light curves.
These Cepheids were later used to trace the shape of the Galactic disk and to provide evidence of the Galactic warp \citep{Chen2019}.
\citet{Yang2018} extracted the WISE light curve to study the variability of red supergiant stars in the Large Magellanic Cloud.
Unlike \citet{Chen2018}, who mainly focuses on the short-timescale variation, \citet{Yang2018} binned the exposures in each visit to one and derived time-series data of eight epochs in 6 years.
We focus on the long-term brightness variation of Be stars and therefore adopted a similar approach as \citet{Yang2018}, but with a much longer time span, that is, 13 years, using the NEOWISE 2023 release.
We describe the extraction of the light curve in section~\ref{sec:wise-lc}.

\subsection{Input catalogues} 
\label{sec:sample}

In order to investigate the light curves of Be stars in the WISE bands and develop an automated method for detecting BeT stars, we first studied a specific subset of Be stars.
This subset consisted of known classical Be stars identified in previous studies, and their sources are listed in the left columns of Table~\ref{tab:samples}. 
Some of these stars are expected to exhibit Be phenomena over the past 13 years, which can be detected from their WISE photometry. 
These light curves were manually identified and classified into BeT stars and non-BeT stars. 
Through this manual classification process, we derived statistical criteria that were subsequently used for automated detection. 
This subset of the sample, consisting of known Be stars, is referred to as the golden sample.
We then applied the criteria determined from the golden sample to a larger scale of early-type stars to detect new BeT stars. 
This sample included the Be stars detected from large-scale surveys, for instance, the Large Sky Area Multi-Object Fibre Spectroscopic Telescope (LAMOST) and The Apache Point Observatory Galactic Evolution Experiment (APOGEE), and the other early-type stars.
We called this sample the extended sample, and the sources are also listed in Table~\ref{tab:samples}.

\begin{table*}
    \caption{Composition of the input catalogue, with the sources and a brief description of the input catalogue. The sources with an asterisk are limited to $\Teff > \SI{7000}{K}$.}
    \centering
    \begin{tabular}{cc}
        \hline\hline
        \multicolumn{2}{c}{The golden sample (CBes only)} \\
        Source & Description \\
        \hline
        \citet{Neiner2011, Neiner2018} & BeSS catalogue\\
        \citet{Lin2015, Wang2022} & LAMOST-based\\
        \citet{Raddi2015} & IPHAS\\
        \citet{Bernhard2018} & ASAS survey\\
        \citet{Granada2018} & WISE (single epoch) \\
        \citet{Vioque2020} & Gaia DR2 \\
        \multicolumn{2}{c}{Total: \num{3510} / \num{2740} (with WISE light curve)} \\ 
        \hline
        \multicolumn{2}{c}{The extended sample} \\
        Source & Description \\
        \hline
        \citet{Roman-Lopes2020} & APOGEE OB stars \\
        \citet{Chojnowski2015} & APOGEE emission line stars \\
        \citet{Zhang-YJ2022, Hou2016} & LAMOST emission line stars\\
        \citet{Xiang2021, Guo2021, Sun2021} & LAMOST OBA stars\\
         \citet{Sprague2022} & APOGEE-Net* \\
         \citet{Majewski2017} & APOGEE DR17* \\
          \citet{Gilmore2022, Randich2022} & Gaia-ESO survey DR5*\\
         \citet{Deng2012, Zhao2012, Liu2014} & LAMOST DR8*\\
         \multicolumn{2}{c}{Total: \num{345571} / \num{202493} (with WISE light curve)} \\
        \hline
    \end{tabular}
    \label{tab:samples}
\end{table*}

\begin{figure*}
    \includegraphics[width=\figwidthl]{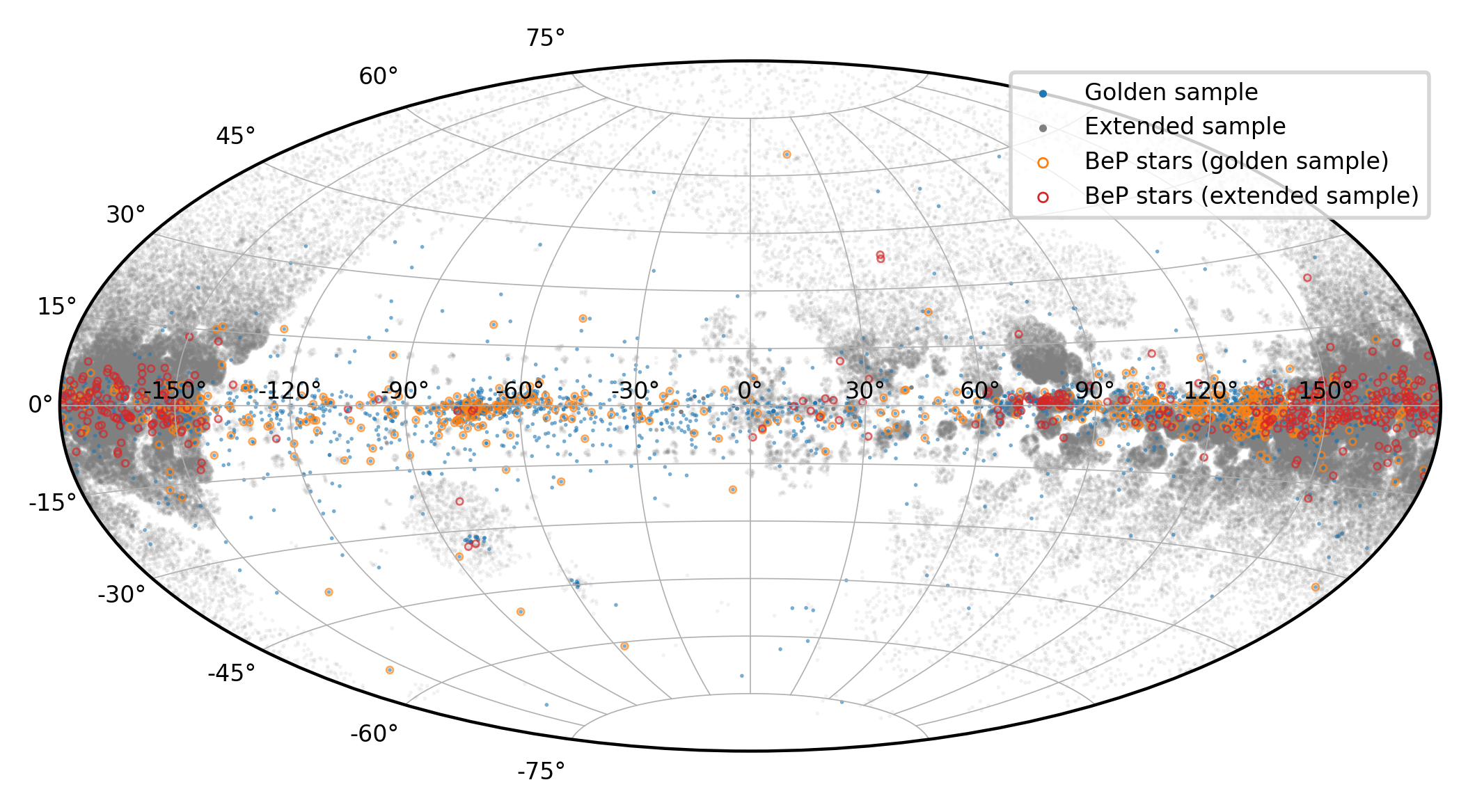}
    \caption{Sky distribution of the input catalogues and the WISE BeT stars. The stars from the golden and extended samples are plotted in blue and grey, and the BeT stars detected from these two samples are circled in orange and red.
    \label{fig:BeT-sky-map}}
\end{figure*}

Figure~\ref{fig:BeT-sky-map} presents the sky positions of the golden and extended sample stars with WISE epoch photometry.
The LAMOST survey provides the dominant entries in $60\degree < l < 210\degree$ (northern hemisphere), while the APOGEE survey is the main contributor in the other area.
We note, however, that only a small number of early-type stars are in the APOGEE survey due to its target selection strategy, and the completeness is much lower in the area with APOGEE coverage alone. 
Future large-scale surveys in the southern hemisphere, such as the galactic surveys in the 4-metre Multi-Object Spectroscopic Telescope (4MOST; see \citealt{de_Jong2019}, \citealt{Bensby2019} and \citealt{Lucatello2023}), will largely fill the southern hemisphere blank in this figure.

\section{Detecting the Be phenomenon in WISE epoch photometry}
\label{sec:wise-lc}

\subsection{WISE binned light curve}

As described in section~\ref{sec:wise-intro}, we adopted an approach similar to the one used in \citet{Yang2018} to extract the WISE light curve.
Because precise $W3$ and $W4$ photometry is only available in the first two years of the WISE mission, we limited our data to the $W1$ and $W2$ bands.
For a given coordinate, we extracted the epoch photometry from the ALLWISE catalogue (with the data in the four-band cryo, three-band cryo, and post-cryo period) and the NEOWISE catalogue within a radius of \SI{3}{''}, corresponding to the angular resolution of these bands.
We then limited the records to those without severe blending, that is, the number of blend components (\texttt{nb}) used in the fit is one, no active blending was implemented (\texttt{na} is 0), and no artefacts (\texttt{cc\_flags} is 0000).
It is known that WISE photometry has a saturation bias when the target is bright, for example, $W1 \lesssim \SI{7}{mag}$ or $W2 \lesssim \SI{6}{mag}$.
We applied the empirical saturation correction provided by WISE\footnote{\url{https://wise2.ipac.caltech.edu/docs/release/neowise/expsup/sec2_1civa.html}} accordingly, and we excluded the exposures with an error in $W1$ or $W2$ larger than \SI{0.05}{mag}.

\begin{figure}
    \includegraphics[width=1\columnwidth]{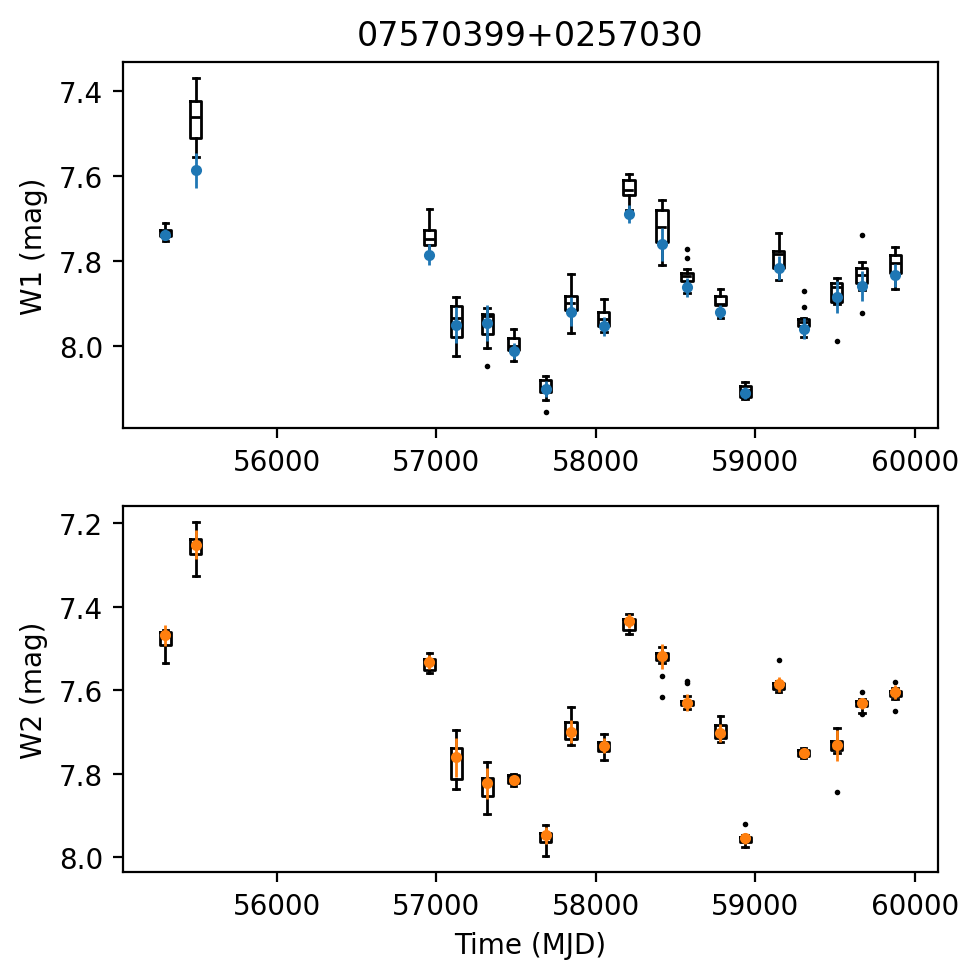}
    \caption{Example of the WISE epoch photometry (box plots with outliers plotted as black dots) and the binned light curve (in blue for $W1$ and orange for $W2$). The saturation correction was performed after the epoch photometry was extracted, and therefore, some part of the binned light curve, especially in $W1$, is lower than the median of the grey points.  
    \label{fig:wise-lc-example}}
\end{figure}

All the epoch photometry entries for each star were then binned by each visit.
The $W1$ and $W2$ magnitude, $(W1-W2)$ colour, and their errors for each visit were calculated as the median and weighted standard deviation of all photometry in the visit. 
Figure~\ref{fig:wise-lc-example} presents an example of the single-epoch photometry and binned light curves in $W1$ and $W2$.
In total, \num{2740} and \num{202493} stars in the golden and extended samples have WISE epoch photometry.
Only the visit-binned photometry was used in the following analysis.

\subsection{Selecting the WISE BeT stars}
\label{sec:selecting-BeT}

\begin{figure}
    \includegraphics[width=1\columnwidth]{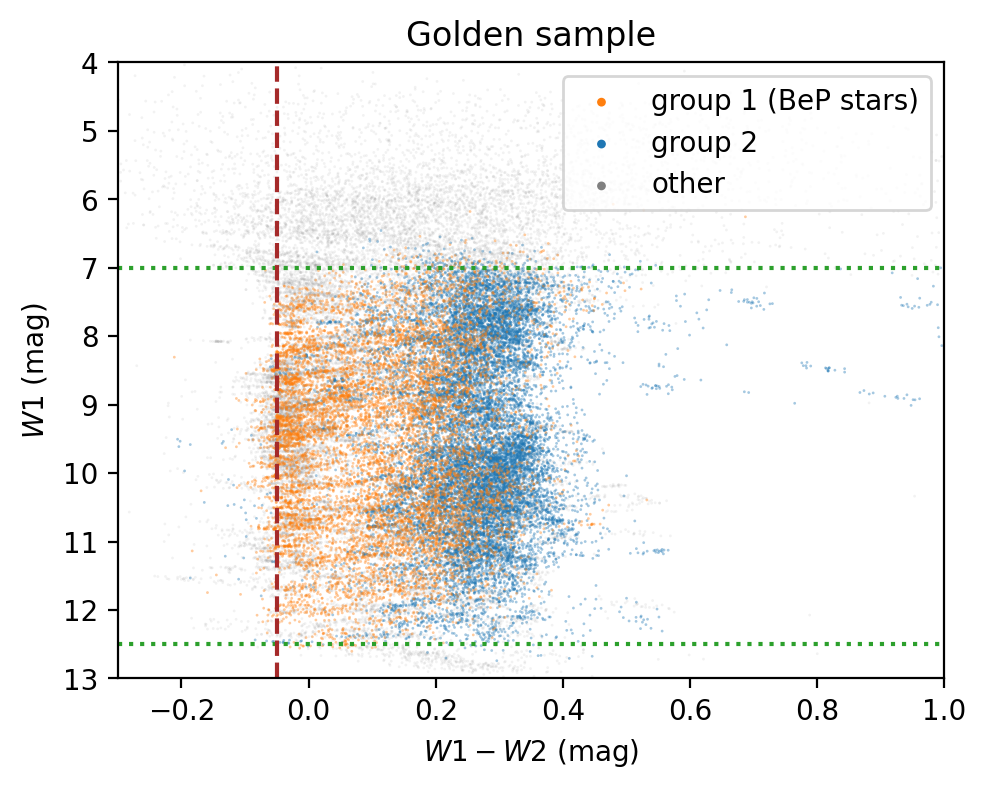}
    \caption{WISE colour-magnitude diagram of the golden sample stars. The vertical brown line indicates the intrinsic colour (\num{-0.05}) of our sample stars, and the horizontal dotted green lines indicate the limit of $W1$ for detections of the BeT stars.
    \label{fig:golden_sample_CMD}}
\end{figure}

\begin{figure}
    \includegraphics[width=1\columnwidth]{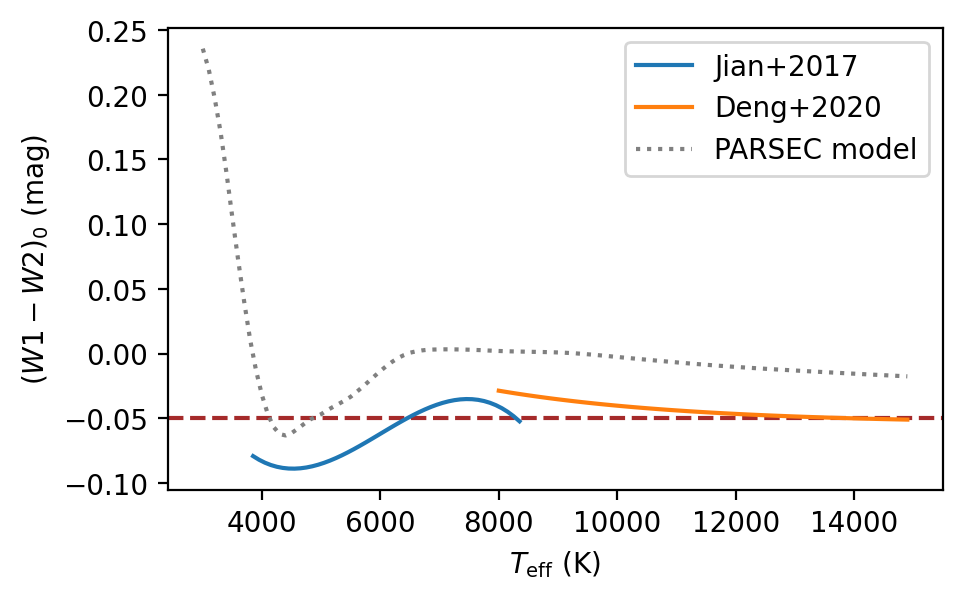}
    \caption{Intrinsic colour $(W1-W2)_0$ vs. $\Teff$ derived from \citet[][blue]{Jian2017}, \citet[][orange]{Deng2020} and PARSEC (dotted grey). The horizontal brown line indicates the color value of \num{-0.05} that was used to select the reddened stars in section~\ref{sec:selecting-BeT}.}
    \label{fig:ic_teff}
\end{figure}

Because \citet{Granada2018} predicted that the Be phenomena will brighten and redden the star, a detection of BeT stars in the WISE bands also relies on these two parameters.
The brightness of the Be stars, for instance, their $W1$ magnitude, is affected by their distance, extinction, and the presence of a disk. It is therefore difficult to detect variation caused by the Be phenomenon using this parameter alone.
The $(W1-W2)$ colour, on the other hand, is critical for detecting the Be phenomenon. 
It is not altered by the distance of the star, and the extinction effect in these bands are similar, that is, $E(W1-W2) = 0.128 A_\mathrm{K_S} \approx 0.0128 A_V$ \citep[][with $A_\mathrm{K_S}$ and $A_\mathrm{V}$ being the extinction in $\mathrm{K_S}$ and $V$ bands]{Xue2016}.
A star with $A_\mathrm{V} = \SI{10}{mag}$ will be reddened by \SI{0.13}{mag} in $(W1-W2)$, which cannot explain the colour distribution in our golden sample (see Figure~\ref{fig:golden_sample_CMD}). 
Further, the intrinsic colour of the central star without the effect of any circumstellar disk, $(W1-W2)_0$, can be treated as a constant for the early-type stars.
Figure~\ref{fig:ic_teff} presents the relation between intrinsic colour and effective temperature ($\Teff$) derived from \citet{Jian2017} and \citet{Deng2020}, calibrated using a large number of stars with stellar parameters derived from the LAMOST and APOGEE surveys.
The intrinsic colour is about $-0.05$\,mag for stars with $\Teff > \SI{6000}{K}$.
The dashed grey curve in the figure represents the $(W1-W2)_0$--$\Teff$ relation from the PARSEC isochrone \citep{Bressan2012}, which is $\sim$\SI{0.05}{mag} systematically redder the other two curves.
Because the model calculation in the infrared would contain some uncertainties due to the presence of numerous molecular bands, we adopted the intrinsic colour from \citet{Jian2017} and \citet{Deng2020}.
The early-type stars in our sample should be centred around $W1-W2 \approx (W1-W2)_0 = \SI{-0.05}{mag}$ in the colour-magnitude diagram if they do not show Be phenomena.
Those showing Be phenomena, on the other hand, would have a colour redder than \SI{-0.05}{mag}.
No reddening correction was performed because a reddened star can still pass the criteria listed below to be a BeT star candidate, and the parameter used in the final classification (the Pearson correlation
coefficient) is not affected by the reddening. 

Figure~\ref{fig:golden_sample_CMD} shows the distribution of the epoch photometry in colour-magnitude diagram (epoch-CMD) for the golden sample stars.
Part of the measurements are centred on the $W1-W2 = \SI{-0.05}{mag}$ vertical line, but many stars have larger colour indices.
These stars would be the BeT star candidates, and we selected the candidate using the following criteria:
\begin{enumerate}
    \item the star has a good photometry quality: with a mean in $W1$ mag, $7 < \overline{W1} < \SI{12.5}{mag}$;
    \item the star is reddened: at least two visits with $W1-W2 > -0.05 + 2 \times \overline{\sigma_{W1 - W2}}$, with $\sigma_{W1 - W2}$ indicating the error in $(W1-W2)$;
    \item $W1$ varied: the peak-to-peak (ptp) range of $W1 > 3\times \overline{\sigma_{W1}}$;
    \item enough measurements: at least five measurements are available.
\end{enumerate}

\begin{figure*}
    \centering
    \includegraphics[width=\figwidthl]{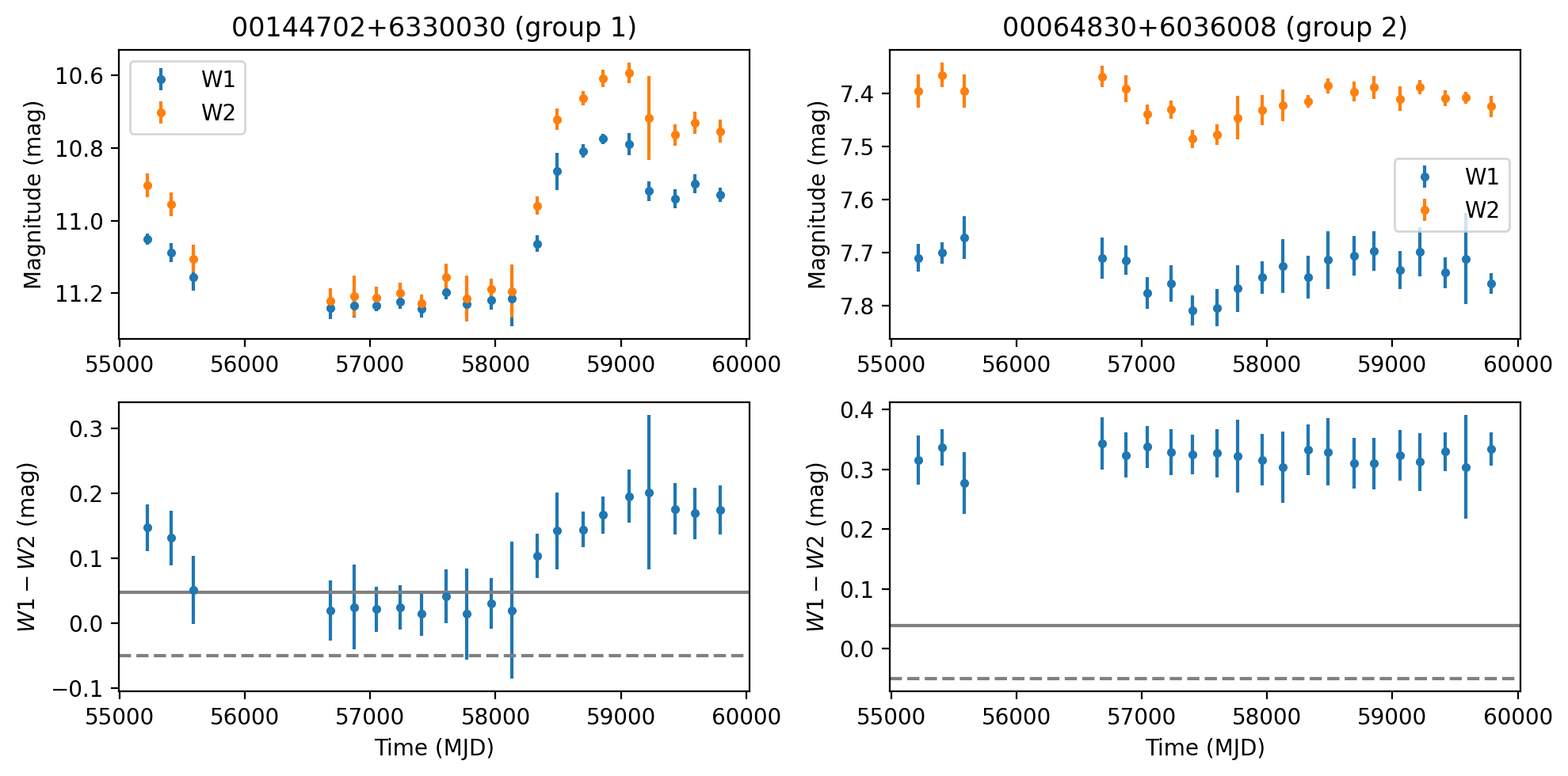}
    \caption{Light-curve examples for group 1 and 2 stars in the golden sample. 
    \label{fig:golden_g12_lc}}
\end{figure*}

These criteria are met by 1340 stars in the golden sample. 
An inspection of their light curves by eye shows that although these stars show some extent of variation in brightness, not all of them show a redder-and-brighter behaviour. 
These stars were then further classified into two groups: group 1, or the WISE BeT stars, with an evident Be phenomenon characteristic (the star becomes brighter or fainter and redder or bluer), and group 2 without this characteristic.
Figure~\ref{fig:golden_g12_lc} presents example light curves for these stars.
The criteria listed above were met by 514 stars, and they were also classified as group 1 stars.

\begin{figure}
    \includegraphics[width=1\columnwidth]{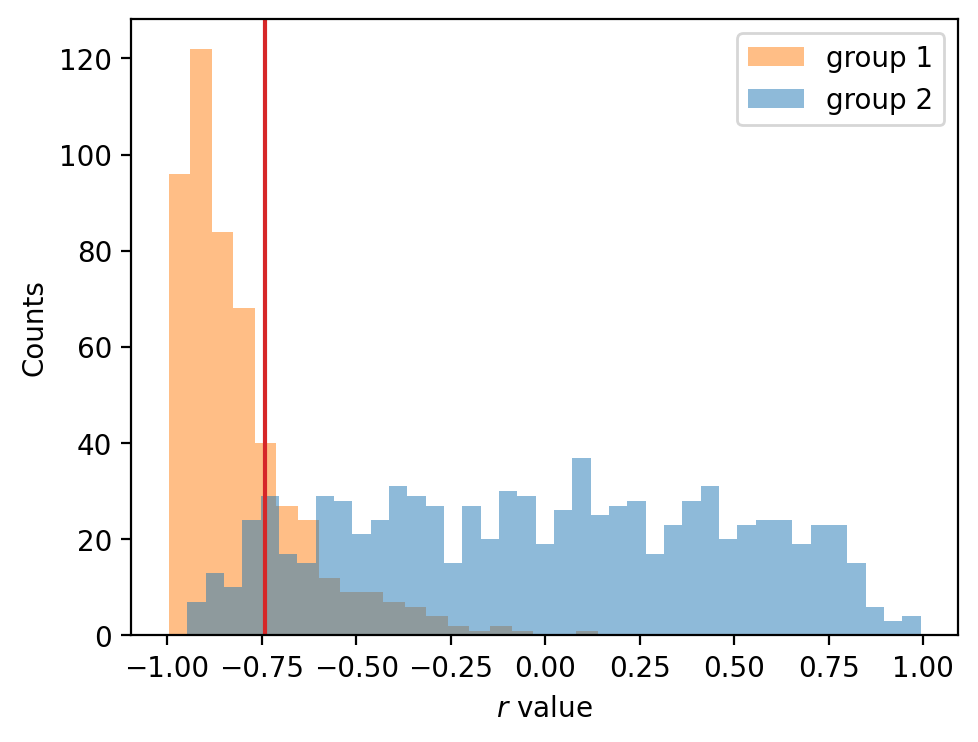}
    \caption{Histogram of $r$ values for group 1 (orange) and 2 (blue) stars. The vertical red line indicates the criteria for classifying group 1 and 2 stars for the extended sample.
    \label{fig:r_value}}
\end{figure}

We then determined which parameter derived from the light curve provided the best separation for group 1 and 2 stars.
Because the colour index is expected to be negatively correlated with the magnitude during the Be phenomenon, the Pearson correlation coefficient, $r$, which describes whether $W1$ and $(W1-W2)$ are correlated, is an ideal parameter.
Figure~\ref{fig:r_value} presents the histogram of the $r$ values, separated by their groups. 
The $r$ value distribution of group\,1 and 2 stars is clearly different: The $r$ values for group\,1 stars are concentrated around $-1$, consistent with the negative correlation between brightness and colour during the Be phenomenon phase, while those for group\,2 stars are similar to the flat distribution.
When we set $r = -0.74$ as the threshold for separating group 1 and 2 stars, we would achieve the same false discovery rate (ratio of the number of group 2 stars with $r < -0.74$ to that of all stars with $r < -0.74$) and false omission rate (ratio of the number of group 1 stars with $r \ge -0.74$ to that of all stars with $r \ge -0.74$) of 13\%.
We then adopted this value as the threshold for separating group\,1 and 2 stars in the extended sample, while keeping all the group\,1 stars in the golden sample in our final BeT catalogue.

\begin{figure}
    \includegraphics[width=1\columnwidth]{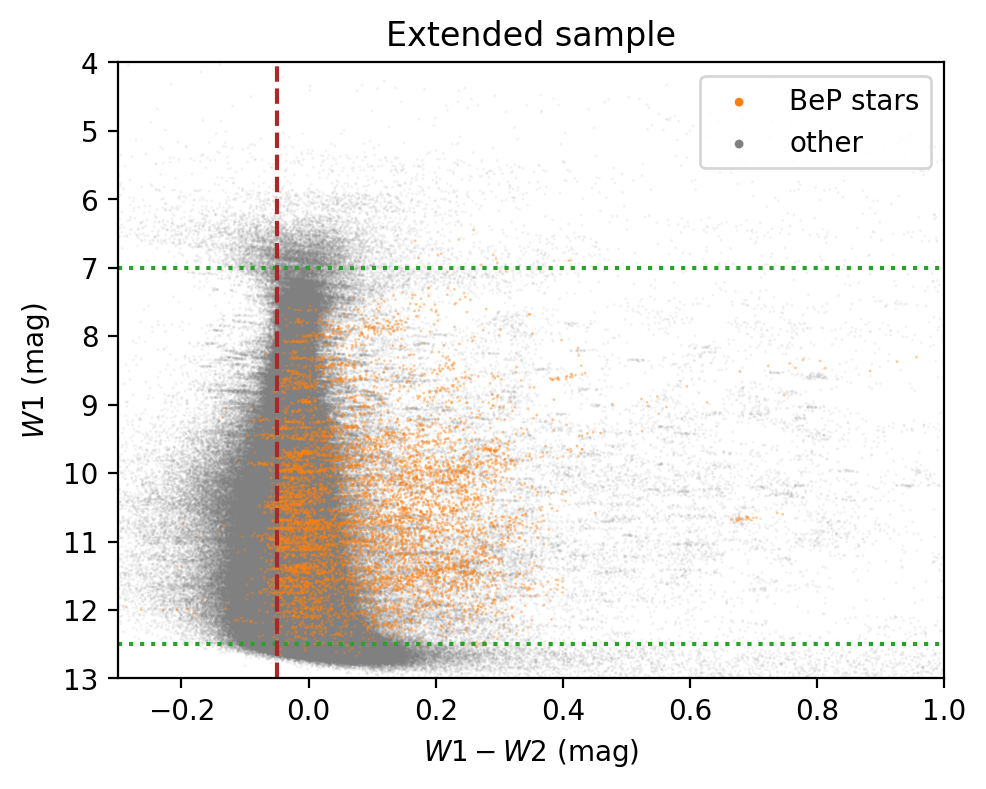}
    \caption{Same as Figure~\ref{fig:golden_sample_CMD}, but for the extended sample. 
    \label{fig:extended_sample_CMD}}
\end{figure}

Figure~\ref{fig:extended_sample_CMD} presents the epoch-CMD of the extended sample.
Because the number of stars in the extended sample is much larger than that in the golden sample, and because most of the stars in the extended sample are not Be stars, the vast majority of the points are located around their intrinsic colour.
Stars with Be phenomena, that is, those with $0.05 < W1-W2 < \SI{0.35}{mag}$, are also obvious. 
Some extent of upper right to lower left pattern is seen in the figure for the BeT stars in Figure~\ref{fig:golden_sample_CMD} as well.

\section{Result}
\label{sec:result}

This section provides the catalogue of stars displaying Be phenomena, their correlation with H$\alpha$ emission, and the behaviour of the outburst and decay phase. 
A total of 916 stars from the golden and extended sample exhibit variations with the characteristic of being redder-and-brighter. They are thereby classified as WISE BeT stars. 
These stars are listed in Table~\ref{tab:bep-catalogue}, and their sky distribution is presented in Figure~\ref{fig:BeT-sky-map}.
Similar to the distribution of OBA stars, the BeT stars are also concentrated around the Galactic disk. 
Of the WISE BeT stars, 145 are assigned a variability type in \citet{Labadie-Bartz2017} or \citet{Bernhard2018}, as shown in Table~\ref{tab:bep-catalogue}, and 116 of these stars are classified to the types that expected to show a Be phenomenon-like variability, including outburst variation (ObV), semi-regular outbursts (SRO), and long-term variation (LTV).
The remaining 736 stars are WISE BeT stars that are newly discovered in this study.

\begin{figure}
    \includegraphics[width=1\columnwidth]{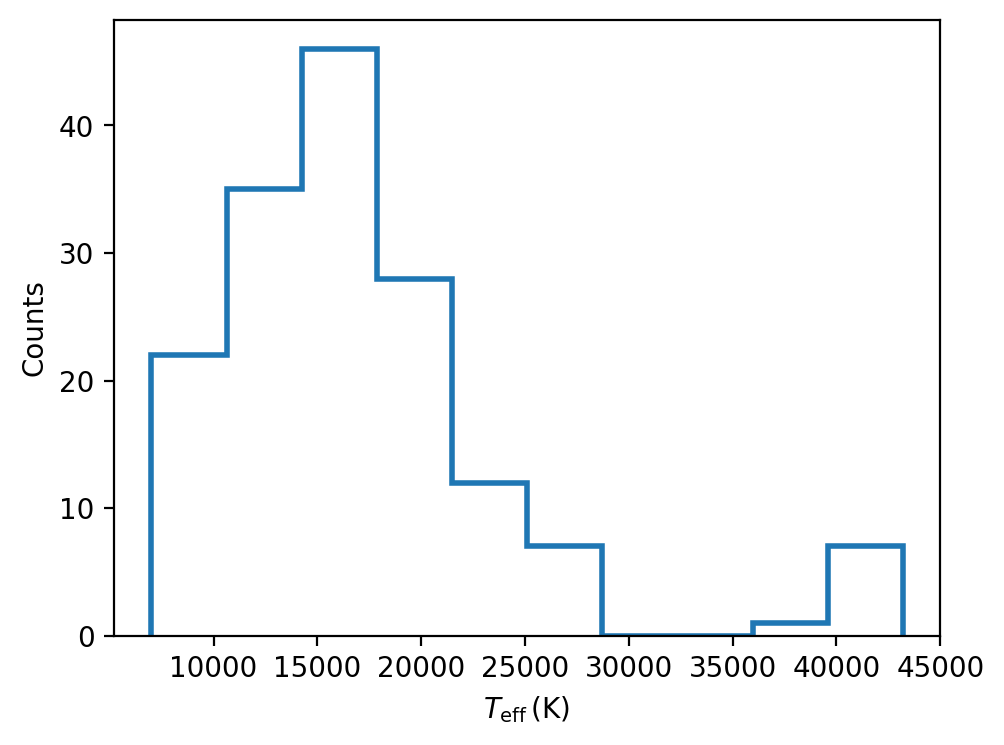}
    \caption{$\Teff$ histogram for BeT stars. 
    \label{fig:bep_teff}}
\end{figure}

\begin{figure}
    \includegraphics[width=1\columnwidth]{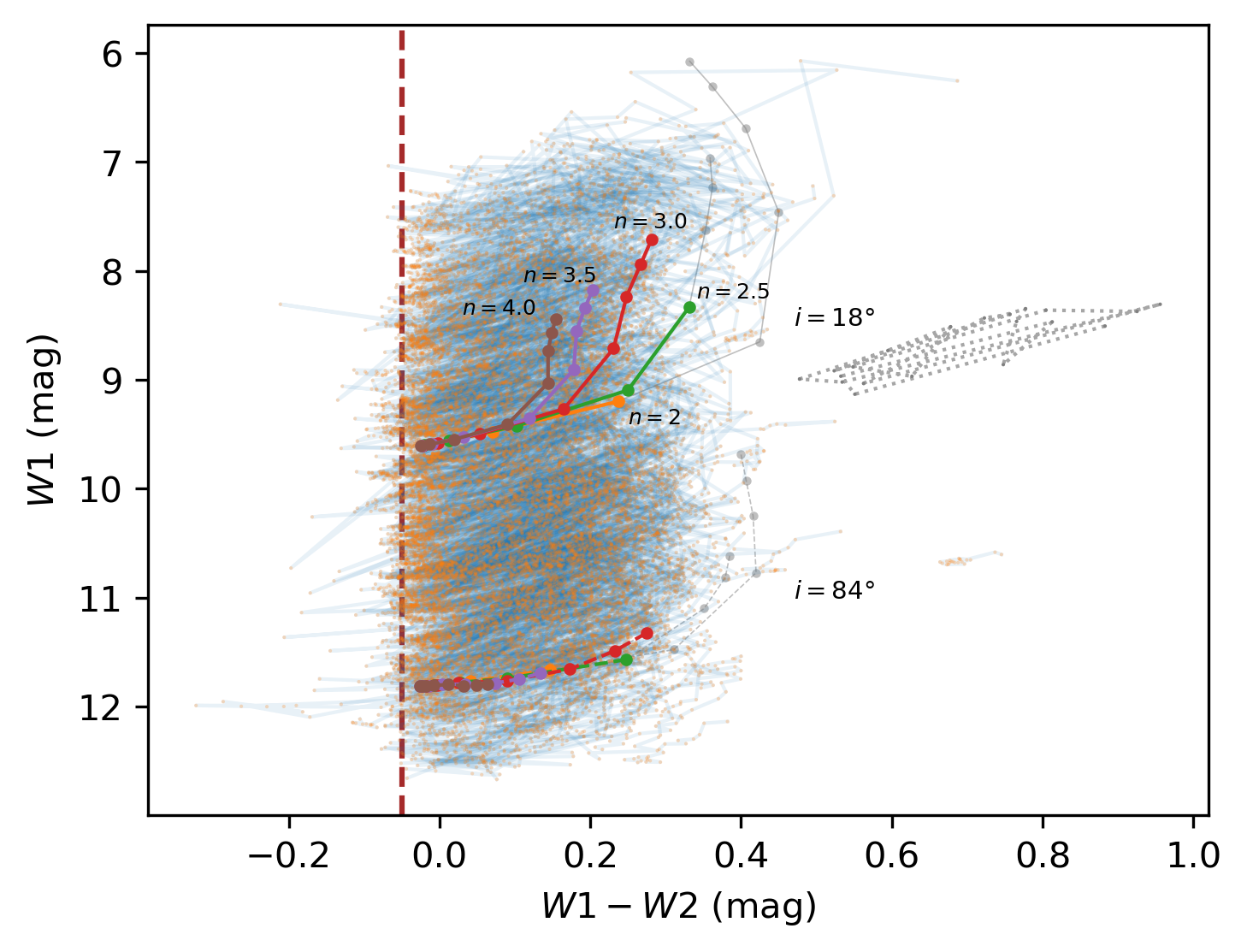}
    \caption{Colour-magnitude diagram of the epoch photometry for the BeT stars. The vertical brown line indicates the intrinsic colour, \SI{-0.05}{mag}, of early-type stars. The star plotted with dotted grey lines is discussed in section~\ref{sec:contamination}. The model prediction from \citet{Granada2018} for B3 type stars is overplotted in two inclination angles ($i=18\degree, 84\degree$), five $n$ and eight $\rho_0$ (increasing from left to right; see the text for more details). The predictions with parameters that are unlikely to be obtained are plotted in grey. $W1$ mag of the predictions is shifted arbitrarily to match the observed magnitude range.
    \label{fig:BeT-CMD}}
\end{figure}

\begin{figure}
    \includegraphics[width=1\columnwidth]{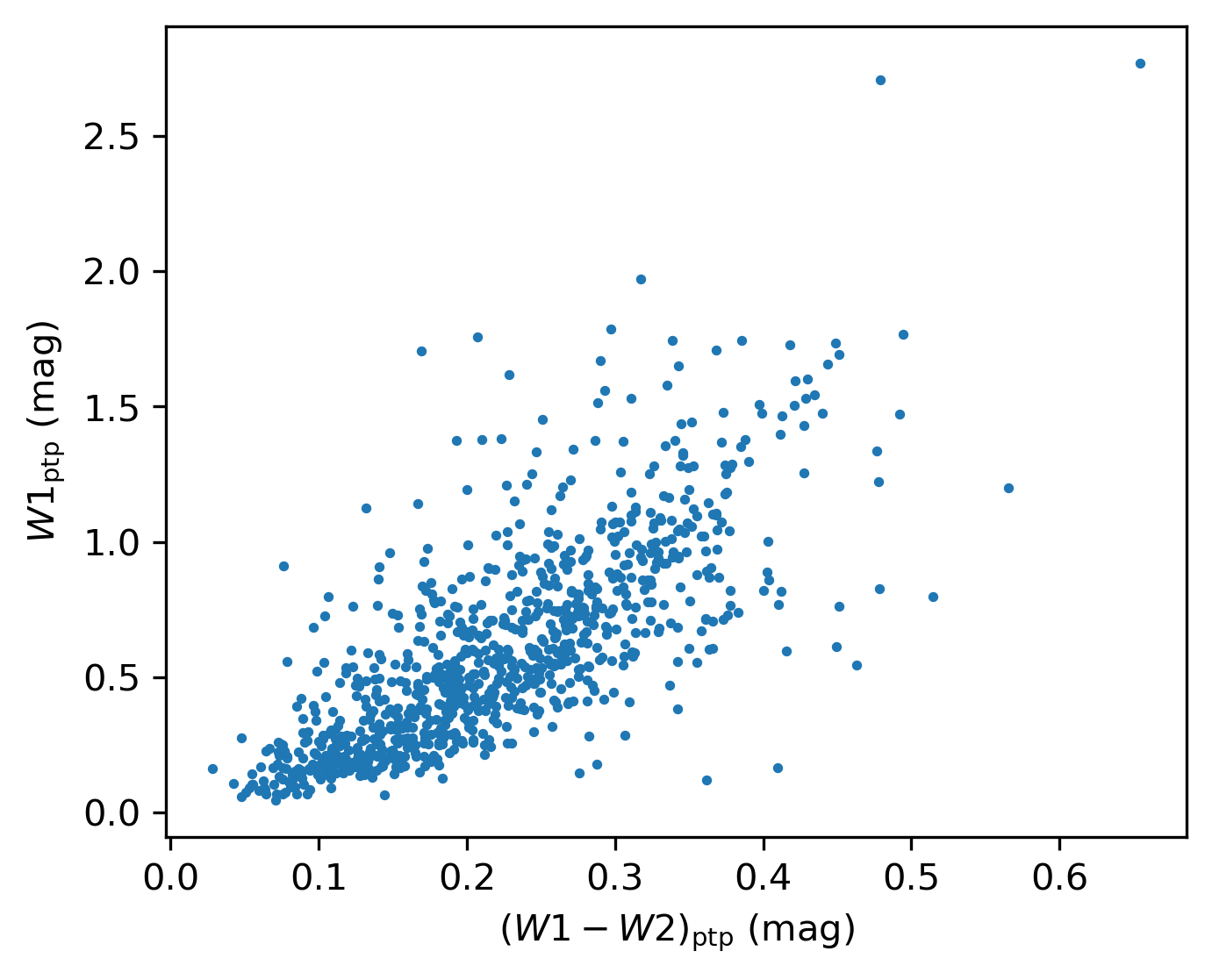}
    \caption{Peak-to-peak variation of $W1$ and $(W1-W2)$ in BeT light curves.
    \label{fig:BeT-ptp}}
\end{figure}

Figure~\ref{fig:bep_teff} displays a histogram of $\Teff$ for the BeT stars. 
The majority of the BeT stars are centred around $\Teff \sim \SI{16000}{K}$, approximately corresponding to B5-type stars. 
However, the distribution also extends to lower $\Teff$, and the sharp cutoff at $\Teff = \SI{10000}{K}$ imposed by the input catalogue criteria suggests the existence of lower-temperature BeT stars.
O-type BeT stars are also present, with $\Teff$ reaching up to $\SI{40000}{K}$.
We note that the $\Teff$ distribution may include some uncertainties, such as the uncertainty of $\Teff$ determined from the spectra for OBA stars (ranging from \num{100} to \SI{750}{K}) and the systematics of the $\Teff$ from a few different sources (ranging from \num{100} to \SI{1700}{K}). 
Figure~\ref{fig:BeT-CMD} presents the epoch-CMD for all WISE BeT stars.
The variation direction, upper right to the lower left, is evident, with a colour range between $-0.05$ and $0.35$\,mag.
The star plotted in grey with a mean $(W1-W2) \sim \SI{0.7}{mag}$  and $W1 \sim \SI{8.6}{mag}$ are discussed in section~\ref{sec:contamination}.
The peak-to-peak ranges of $W1$ and $(W1-W2)$ for all BeT stars are computed and presented in Table~\ref{tab:bep-catalogue}.
The distribution of these two values is shown in Figure~\ref{fig:BeT-ptp}, with most of the $W1_\mathrm{ptp}$ within \SI{1.5}{mag} and $(W1-W2)_\mathrm{ptp}$ within \SI{0.4}{mag}.
Notably, a positive correlation is observed between these two parameters, indicating that these variations are likely to originate from the same type of outburst event.

\begin{sidewaystable*}
\caption{First 10 rows for the BeT stars selected from the golden and extended sample, with their coordinates (in J2000), name, the peak-to-peak range of $(W1-W2)$ and $W1$, the Pearson correlation coefficient ($r$), the mean time span of the outburst ($\overline{t_\mathrm{O}}$) and decay ($\overline{t_\mathrm{D}}$), the sample source, the open cluster to which it belongs, the variable type in \citet{Labadie-Bartz2017} or \citet{Bernhard2018} (TypeL), and the object type in Simbad database (TypeS). The whole table is available at the CDS.}
\centering
    \begin{tabular}{cccccccccccc}
    \hline\hline
        RA &   DEC & Name & $(W1-W2)_\mathrm{ptp}$ & $W1_\mathrm{ptp}$ & $r$ & $\overline{t_\mathrm{O}}$ & $\overline{t_\mathrm{D}}$ & Sample & Cluster & TypeL & TypeS\\
        (\degree) & (\degree) &  & (mag) & (mag) &  & (day) & (day) & & \\
        \hline
         $0.147200$ & $62.554900$ & 00003533+6233176 &    $0.20$ & $0.27$ & $-0.97$ &               ... &               ... &   golden &      ... &    ... &   Star \\
         $2.108293$ & $61.877024$ & 00082599+6152373 &    $0.35$ & $1.28$ & $-0.81$ &           $1010$ &            $758$ &   golden &      ... &    ... &    Em* \\
         $2.695500$ & $63.172936$ & 00104692+6310226 &    $0.14$ & $0.16$ & $-0.81$ &               ... &               ... &   golden &      ... &    ... &    Be* \\
         $2.839740$ & $64.460266$ & 00112154+6427370 &    $0.23$ & $0.92$ & $-0.72$ &            $389$ &            $362$ &   golden &      ... &    ... &     ... \\
         $3.114733$ & $62.978124$ & 00122754+6258412 &    $0.15$ & $0.74$ & $-0.82$ &           $1415$ &               ... &   golden &      ... &    ... &    Em* \\
         $3.351400$ & $59.220400$ & 00132434+5913134 &    $0.27$ & $0.81$ & $-0.82$ &           $1318$ &            $661$ &   golden &      ... &    ... &    Em* \\
         $3.502890$ & $61.093407$ & 00140069+6105363 &    $0.15$ & $0.31$ & $-0.86$ &               ... &               ... &   golden &      ... &    ... &     ... \\
         $3.695937$ & $63.500830$ & 00144702+6330030 &    $0.19$ & $0.47$ & $-0.92$ &           $1579$ &               ... &   golden &      ... &    ... &   Star \\
         $6.621700$ & $60.024400$ & 00262921+6001278 &    $0.13$ & $1.12$ & $-0.83$ &           $1441$ &            $753$ &   golden &      ... &    ... & PulsV* \\
         $7.373112$ & $60.289947$ & 00292955+6017238 &    $0.40$ & $1.00$ & $-0.96$ &               ... &               ... &   golden &      ... &    ... &    Em* \\
         $7.397900$ & $60.304900$ & 00293550+6018176 &    $0.27$ & $0.77$ & $-0.97$ &           $1260$ &               ... &   golden &      ... &    ... &    Em* \\
         $0.970200$ & $57.597200$ & 00035285+5735499 &    $0.31$ & $0.60$ & $-0.96$ &            $377$ &            $458$ & extended &      ... &    ... &     ... \\
         $5.932200$ & $59.848100$ & 00234373+5950532 &    $0.32$ & $0.86$ & $-0.98$ &               ... &               ... & extended &      ... &    ... &     V* \\
         $8.298350$ & $51.668610$ & 00331160+5140070 &    $0.10$ & $0.52$ & $-0.88$ &               ... &               ... & extended &      ... &    ... &    Be* \\
         $8.640900$ & $57.679400$ & 00343382+5740458 &    $0.13$ & $0.19$ & $-0.95$ &               ... &               ... & extended &      ... &    ... &   Star \\
        $16.114240$ & $57.940650$ & 01042742+5756263 &    $0.26$ & $0.70$ & $-0.94$ &               ... &               ... & extended &      ... &    ... &    Be* \\
        $21.902000$ & $59.096600$ & 01273648+5905478 &    $0.33$ & $0.96$ & $-0.85$ &            $327$ &            $476$ & extended &      ... &    ... &   Star \\
        $23.988940$ & $58.153580$ & 01355735+5809129 &    $0.27$ & $0.63$ & $-0.89$ &               ... &               ... & extended &      ... &    ... &    Be* \\
        $30.442792$ & $61.752712$ & 02014627+6145098 &    $0.10$ & $0.25$ & $-0.93$ &               ... &               ... & extended &      ... &    ... &   Star \\
        $31.214399$ & $60.215000$ & 02045146+6012540 &    $0.17$ & $0.50$ & $-0.90$ &            $397$ &            $575$ & extended &      ... &    ... &     ... \\
        $31.343300$ & $60.194900$ & 02052239+6011416 &    $0.15$ & $0.27$ & $-0.96$ &               ... &               ... & extended &      ... &    ... &   Star \\
        $33.992061$ & $57.006676$ & 02155809+5700240 &    $0.10$ & $0.40$ & $-0.91$ &            $519$ &            $376$ & extended &      ... &    ... &    Em* \\
    \hline
    \end{tabular}
    
    \label{tab:bep-catalogue}
\end{sidewaystable*}

\subsection{Correlation between $(W1-W2)$ and H$\alpha$ emission}

\begin{figure}
    \includegraphics[width=1\columnwidth]{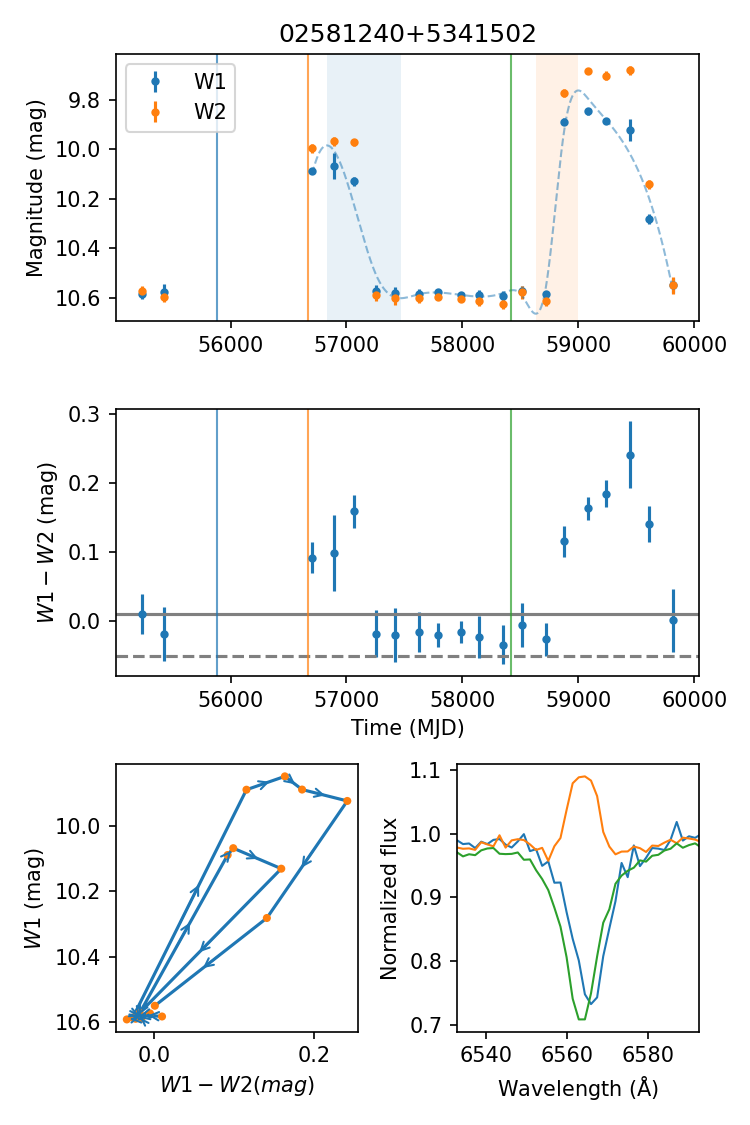}
    \caption{Example of a WISE light curve for the $W1$ magnitude (upper panel) and $(W1-W2)$ colour (middle panel). The CMD is similar to that in Figure~\ref{fig:BeT-CMD}, but is plotted only for this star in the lower left panel, and the H$\alpha$ spectra from LAMOST DR8 are plotted in the lower right panel. The vertical light blue and light orange strips in the upper panel indicate the classified outburst and decay phase in the light curve (section~\ref{sec:od}). All the figures of the light curve are available online.}
    \label{fig:lc_lasmost_example}
\end{figure}

\begin{figure}
    \includegraphics[width=1\columnwidth]{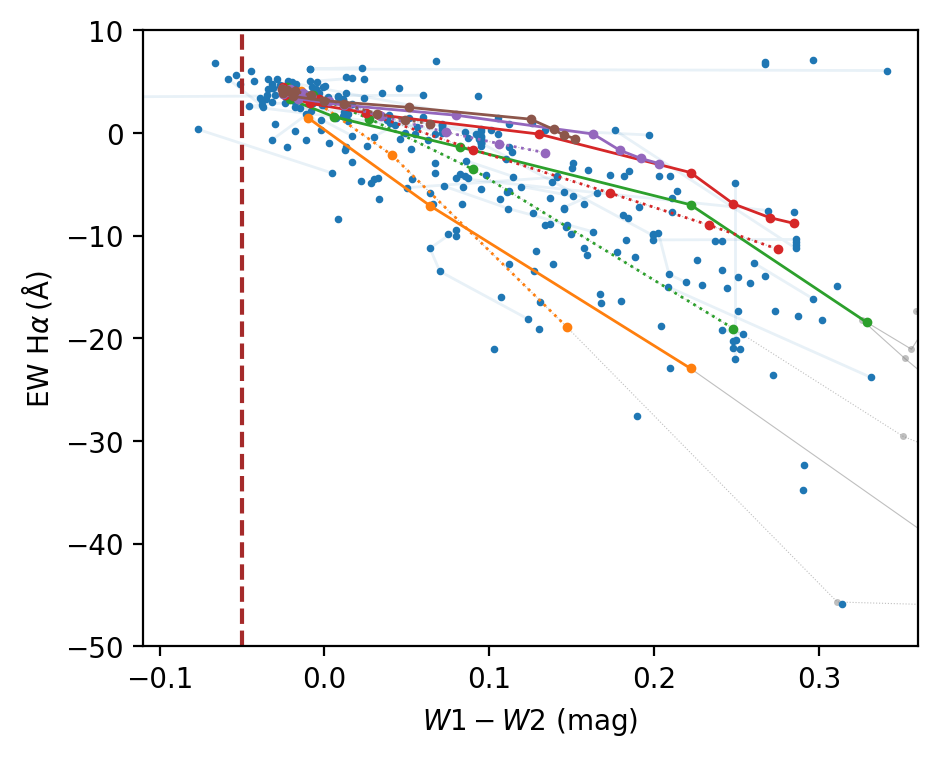}
    \caption{$(W1-W2)$ colour vs. the $EW$ of the H$\alpha$ line. The stars with multiple epoch measurements are connected with light blue lines. The model prediction from \citet{Granada2018} is overplotted, and the colour and line style follow those in Figure~\ref{fig:BeT-CMD}.
    \label{fig:BeT-Ha}}
\end{figure}

The decretion disk during the Be phenomenon is expected to result in the emission of certain spectral lines, such as the H$\alpha$ line \citep{Rivinius2013}. 
The LAMOST survey covers the wavelength range of the H$\alpha$ spectral line, thus it is an ideal dataset on which to confirm whether the variation in WISE bands also correlates with the H$\alpha$ emission.
We cross-matched the catalogue of our BeT stars with the LAMOST DR8 low-resolution catalogue in a radius of $3''$. 
There are 574 spectra from 368 BeT stars with LAMOST observation, as listed in Table~\ref{tab:lamost-catalogue}. 
Time-series spectra are available for some of the stars, for instance, star 02581240+5341502, as shown in Figure~\ref{fig:lc_lasmost_example}.
The first (blue) and third (green) LAMOST observation for this star are clearly in a phase close to $W1-W2=\SI{-0.05}{mag}$ (without a decretion disk), and the second spectra (orange) were obtained when its colour is redder (with a decretion disk).
The corresponding H$\alpha$ emission is stronger for the second spectra, that is, when the stellar colour exceeds \SI{-0.05}{mag}.
This correlation suggests that the brightness variation is indeed caused by disk events.
We then quantified the strengths of the H$\alpha$ line in BeT stars using the equivalent width ($EW$) of the pixels within \SI{20}{\angstrom} of its centre wavelength (\SI{6562.8}{\angstrom}), with a positive $EW$ representing absorption.
Figure~\ref{fig:BeT-Ha} presents the correlation between $(W1-W2)$ colour and H$\alpha$ $EW$.
The colour was taken from the closest WISE observation when the LAMOST spectra were obtained within 180 days.
Most of the H$\alpha$ lines are in absorption when the colour is close to \SI{-0.05}{mag}, and as the stars become redder, the line gradually becomes emission and shows a negative relation with $(W1-W2)$.
This correlation again states that the variation we detected is very likely caused by the Be phenomenon, and whether a star has a $(W1-W2)$ larger than \SI{-0.05}{mag} can be used as the criterion for the presence of a decretion disk. 

\begin{table*}
    \caption{First ten rows of the LAMOST spectra of BeT stars, with the name, observation date (lmjd), plan id (planid), spectrograph id (spid), fiber id (fiberid), closest match $(W1-W2)$ and the $EW$ of the H$\alpha$ line of the spectra. The whole table is available at the CDS.}
    \centering
    \begin{tabular}{ccccccc}
    \hline\hline
      name &  lmjd &  planid &  spid &  fiberid & $W1-W2$ & $EW$ H$\alpha$  \\
       & & & & & (mag) & (\si{\angstrom}) \\
    \hline
    00035285+5735499 &          56982 & HD000408N565515V01 &             15 &               104 &  $0.09$ &  $-4$ \\
    00343382+5740458 &          56592 & HD003103N554209V01 &             11 &               122 & $-0.03$ &   $7$ \\
    00571666+5836557 &          57651 & HD005305N571308V01 &              9 &                77 & $-0.02$ &   $4$ \\
    01042742+5756263 &          57651 & HD005305N571308V01 &             13 &               168 &  $0.23$ & $-15$ \\
    01560714+5633158 &          57328 & HD020325N544136V01 &             16 &               132 & $-0.05$ &    ... \\
    02045146+6012540 &          55875 &              B7505 &             16 &                59 &      ... &    ... \\
    02045146+6012540 &          55910 &             B91003 &             16 &                59 &      ... &   $4$ \\
    02085750+5621086 &          57328 & HD020325N544136V01 &             12 &               150 &  $0.11$ &  $-6$ \\
    02130051+5506396 &          57328 & HD020325N544136V01 &              9 &               135 &  $0.11$ &  $-1$ \\
    02162132+5525358 &          57328 & HD020325N544136V01 &             13 &               139 &  $0.24$ & $-13$ \\
    \hline
    \end{tabular}
    \label{tab:lamost-catalogue}
\end{table*}

\begin{figure*}
        \centering
        \subfloat{\includegraphics[width=\figwidths]{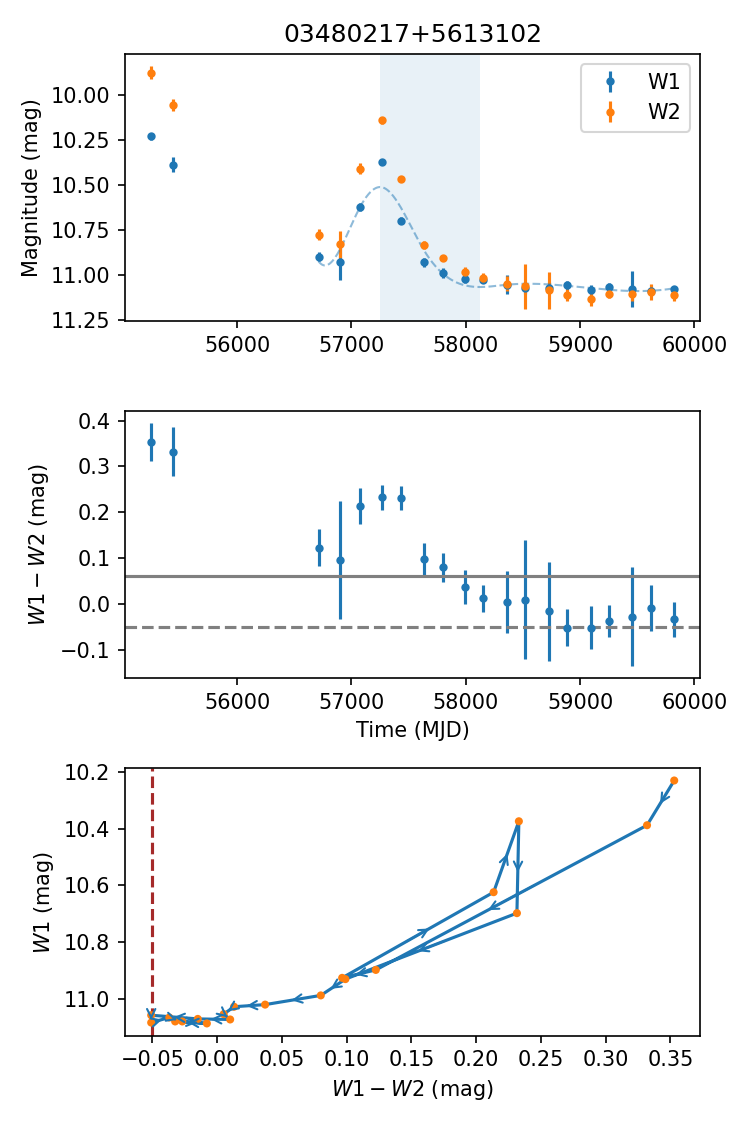}}\hspace{5pt}
        \subfloat{\includegraphics[width=\figwidths]{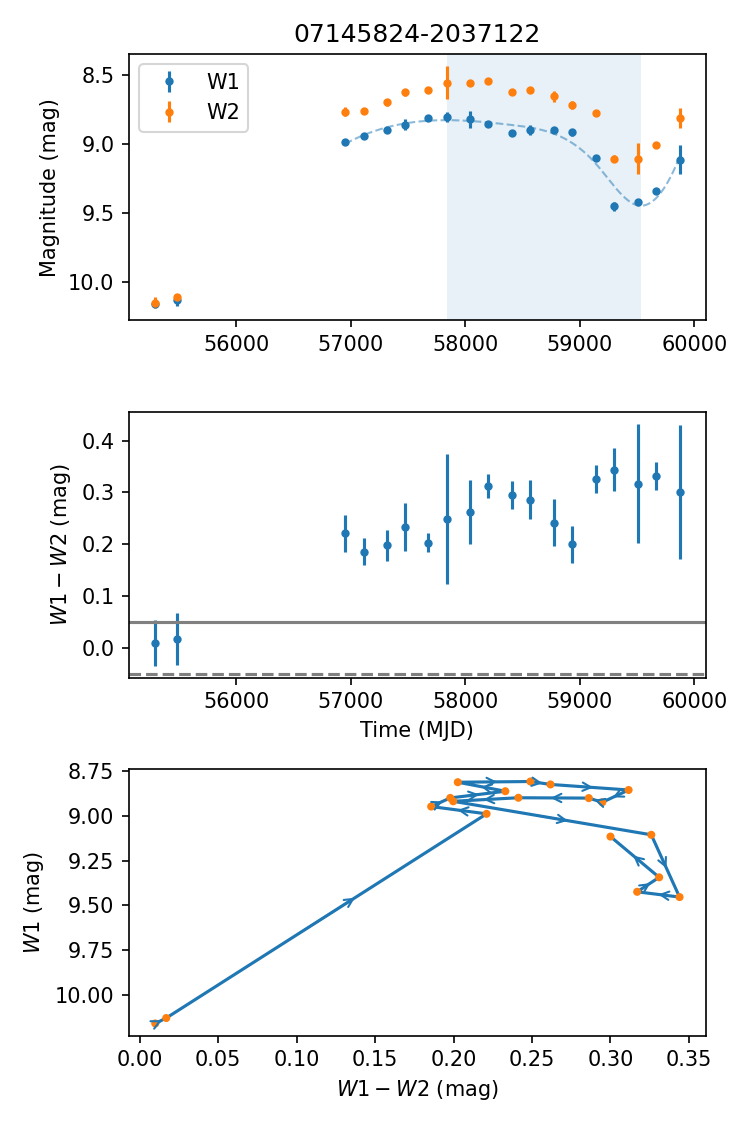}}\\
        \subfloat{\includegraphics[width=\figwidths]{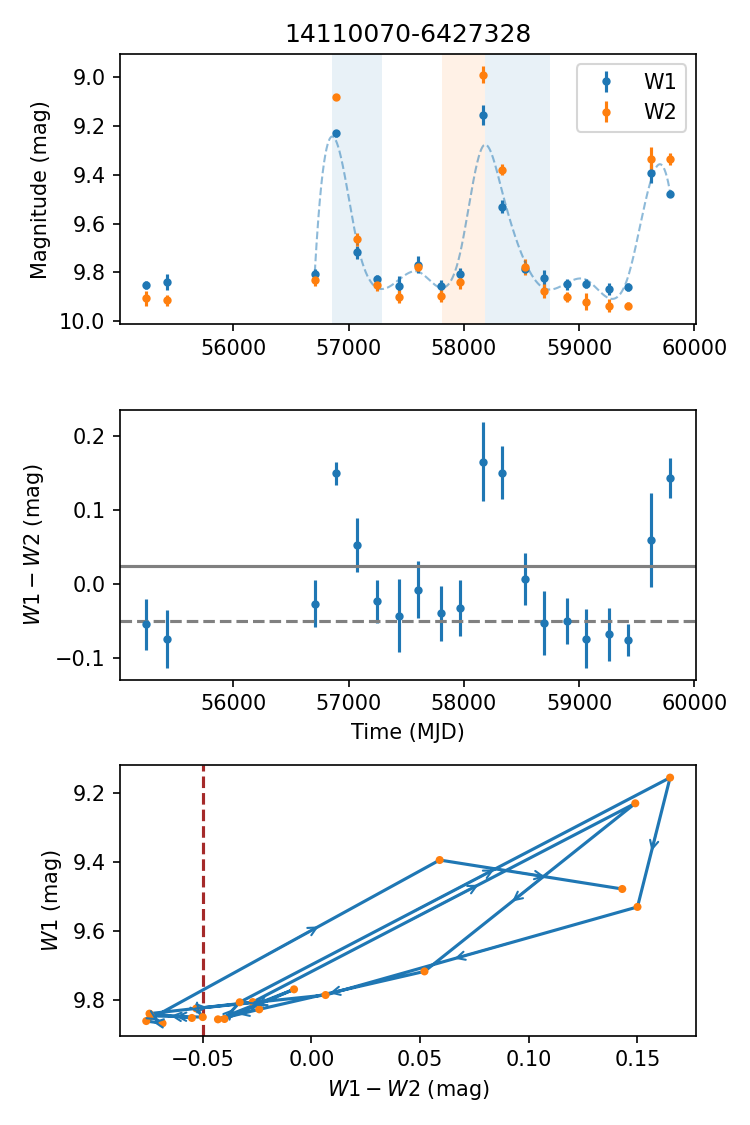}}\hspace{5pt}
        \subfloat{\includegraphics[width=\figwidths]{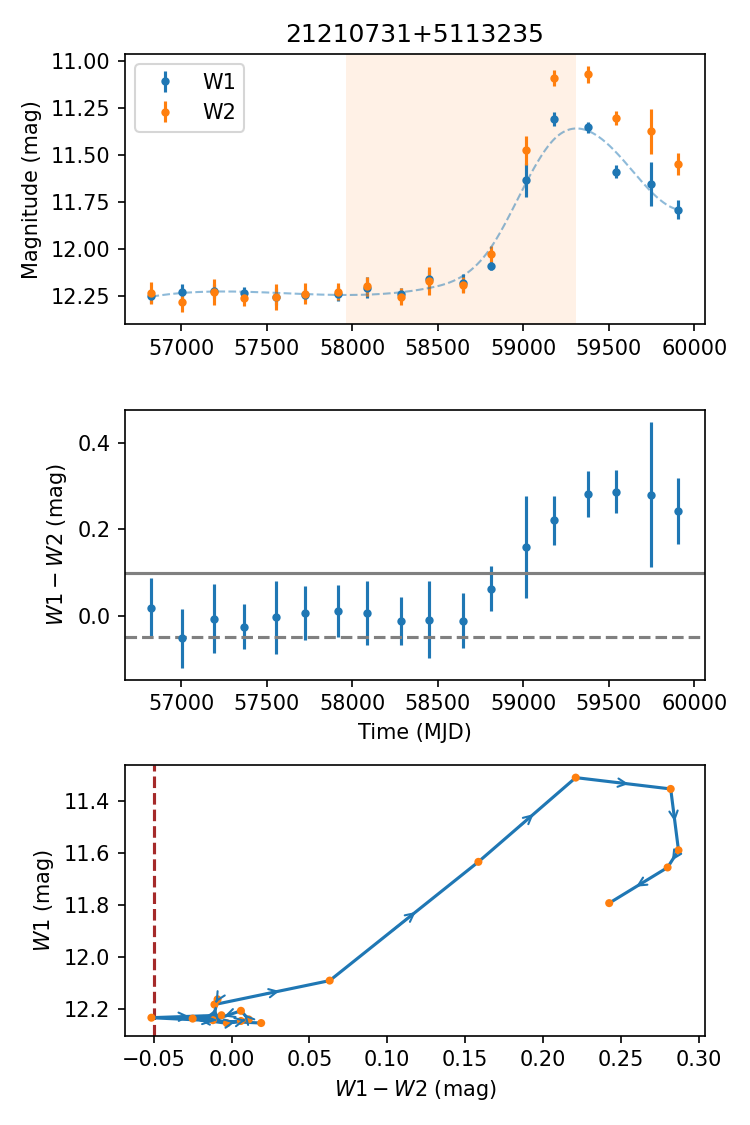}}
        \caption{Same as Figure~\ref{fig:lc_lasmost_example}, but for randomly selected BeT stars from the golden sample. All the figures of the light curve are available online.}
    \label{fig:lc_golden_example}
\end{figure*}

\begin{figure*}
	\centering
	\subfloat{\includegraphics[width=\figwidths]{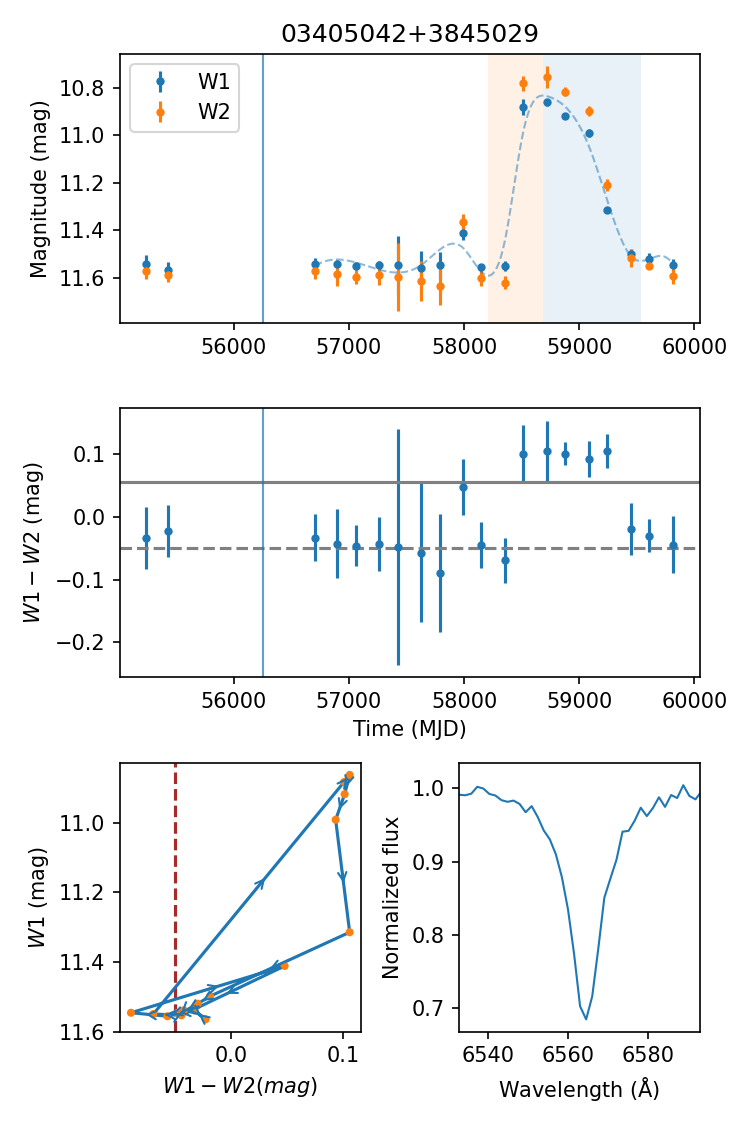}}\hspace{5pt}
	\subfloat{\includegraphics[width=\figwidths]{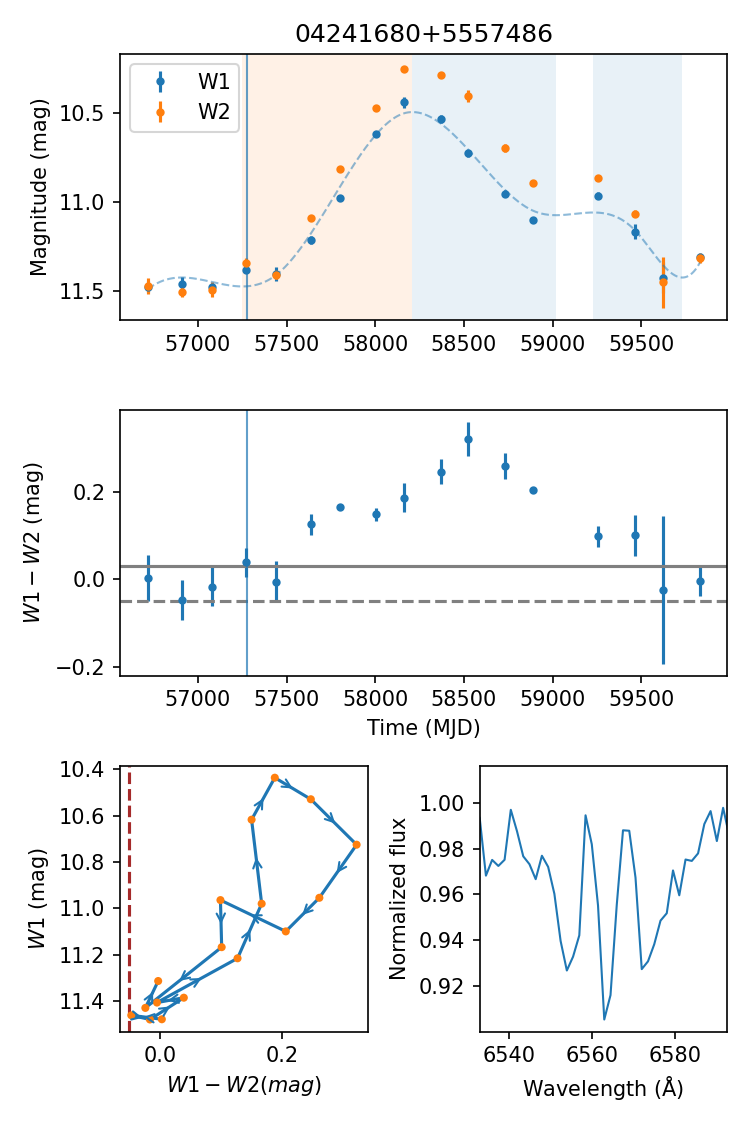}}\\
	\subfloat{\includegraphics[width=\figwidths]{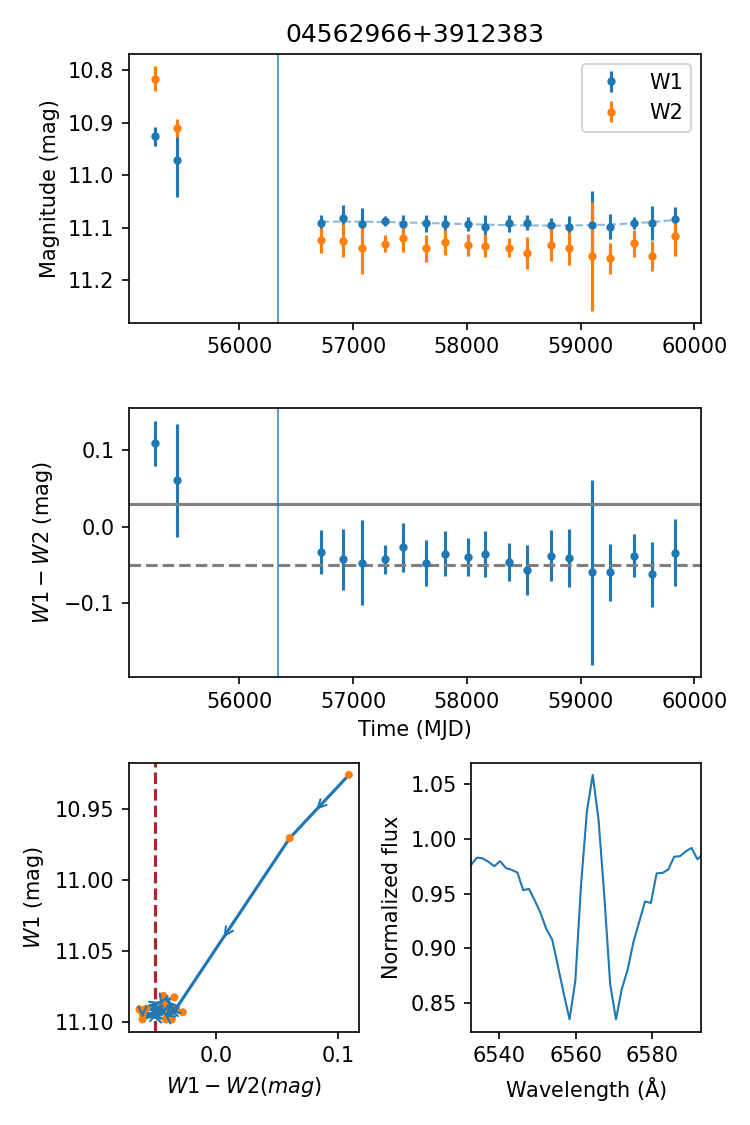}}\hspace{5pt}
	\subfloat{\includegraphics[width=\figwidths]{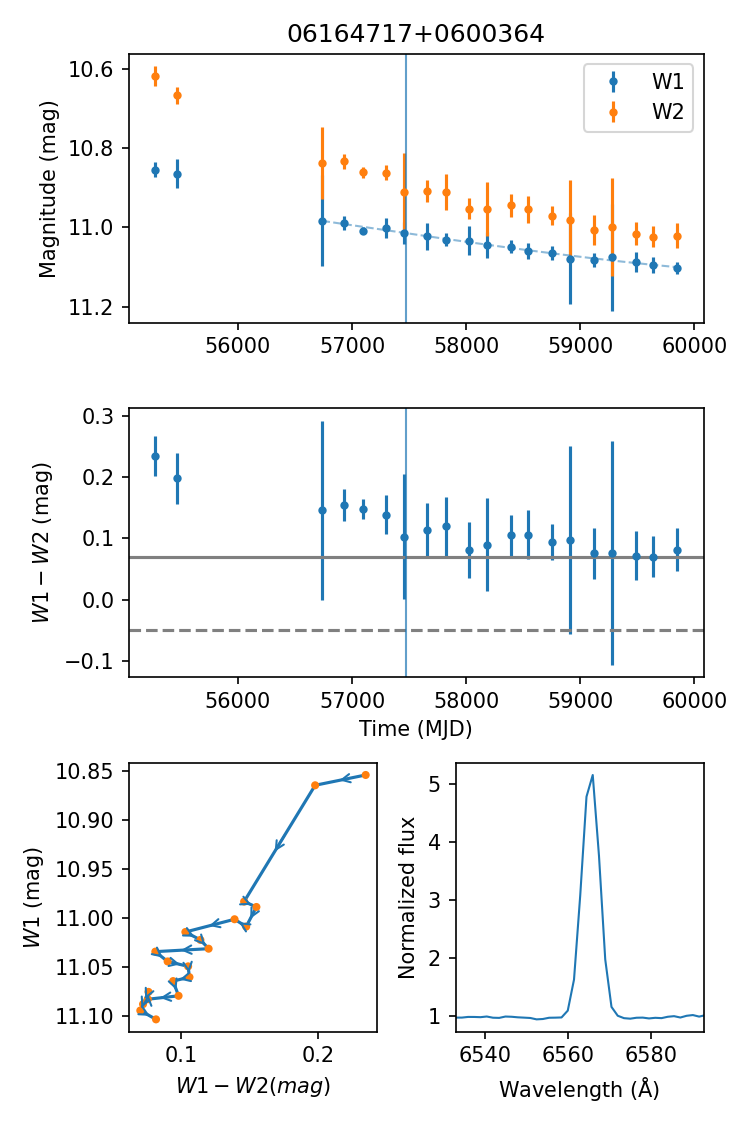}}
        \caption{Same as Figure~\ref{fig:lc_lasmost_example}, but for randomly selected BeT stars from the extended sample. All the figures of the light curve are available online.}
    \label{fig:lc_extended_example}
\end{figure*}

\subsection{Duration and behaviour of the outburst and decay phase}
\label{sec:od}

\begin{figure}
    \includegraphics[width=1\columnwidth]{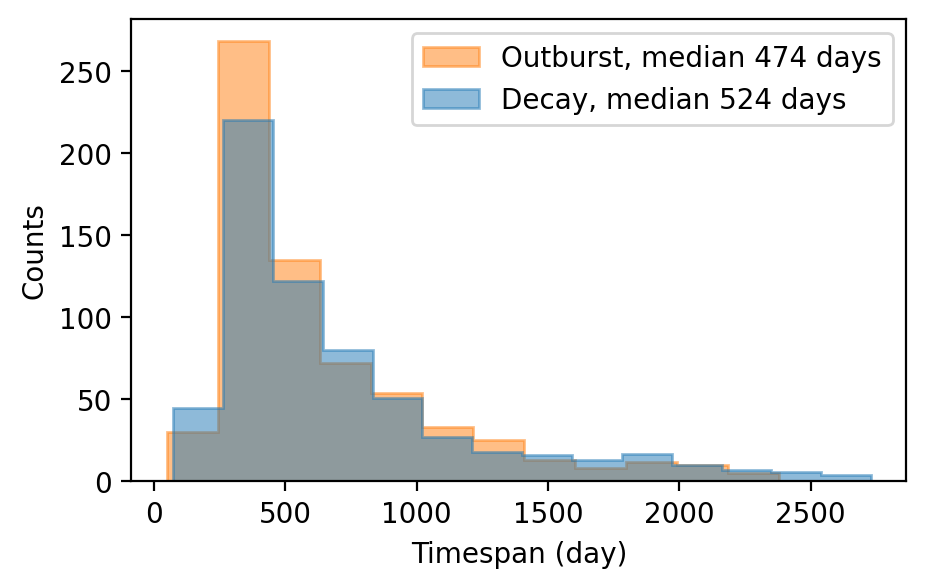}
    \caption{Histogram of the outburst and decay time span.
    \label{fig:OD-timespan}}
\end{figure}

\begin{figure}
    \includegraphics[width=1\columnwidth]{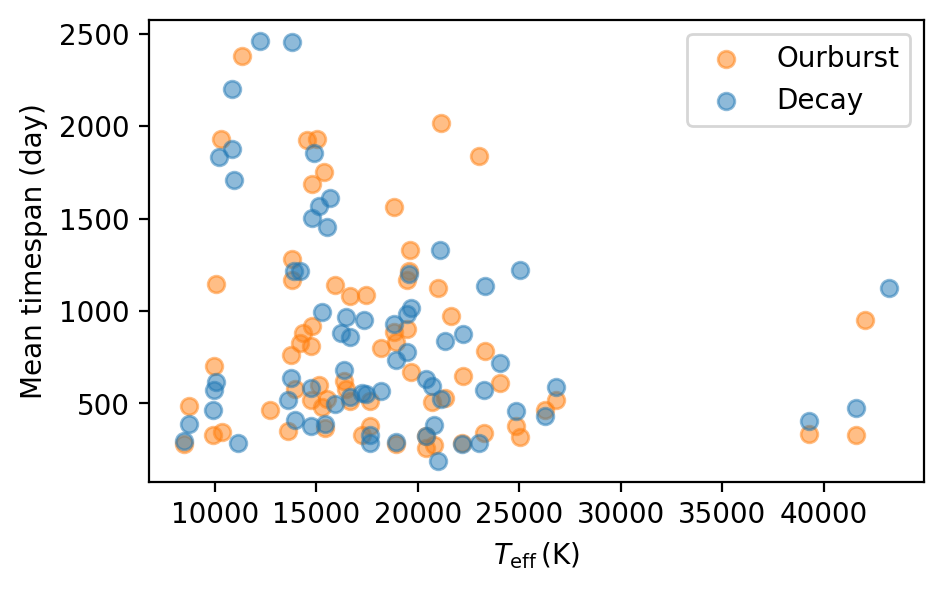}
    \caption{Mean time span of the outburst ($\overline{t_\mathrm{O}}$) and decay phase ($\overline{t_\mathrm{D}}$) for each star vs. $\Teff$.
    \label{fig:OD-Teff}}
\end{figure}

The duration of the disk build-up and dissipation phase of Be phenomena is of interest because it may provide constraints on the mechanism that triggers and maintains the disk, and it also provides guidance about how the future time-series observation on Be phenomena, whether photometry or spectroscopy, should be designed.
We used a spline function to fit the $W1$ light curve of the BeT stars and defined the brightness increase and decrease events with $\Delta W1$ larger than $0.1$\,mag as O(utburst) or D(ecay) phase. 
The phases reaching the start or end of the light curve or those covering the three-year gap of the observation were excluded because the time span of these phases has large uncertainties.
Figure~\ref{fig:lc_golden_example} and \ref{fig:lc_extended_example} show examples for the phase classification.
The orange periods are defined as the O phases, and the blue periods are defined as the D phases.
We note that the spline function was only used as phase definition and does not reflect the actual amount of variation in $W1$.
The duration of all the O and D phases in all BeT stars is shown in Figure~\ref{fig:OD-timespan}.
The distributions of the O and D phases are similar, with a peak at about 300 days and a long tail towards a longer duration. 
The mean O/D time spans for each star are presented in Table~\ref{tab:bep-catalogue}.
The mean time spans cover a wide range for stars with $\Teff \sim \SI{15000}{K}$, as shown in Figure~\ref{fig:OD-Teff}, while those with higher temperatures are mostly within \SI{1000}{days}.
For 323 BeT stars with the mean O and D time span measured, 60\% (195) have a decay time span longer than the outburst.
This is consistent with the conclusion from the VDD model \citep{Haubois2012}.
After the binning, the WISE light curve cannot detect brightness variation shorter than $\approx 180$\,days, and therefore, the duration distribution in 0--180 days is incomplete.
The median values of all the O and D phases are 474 and 524 days, suggesting that a baseline longer than these values would be necessary if we wished to detect new Be phenomena. 

The behaviour of the outburst and decay phase is slightly different.
As shown in the CMDs in Figure~\ref{fig:lc_lasmost_example}, \ref{fig:lc_golden_example}, and \ref{fig:lc_extended_example}, the trajectory of the outburst and decay phase, when well sampled, form a clockwise loop.
The increase in brightness precedes the rise in the colour index, resulting in a steeper trajectory during the outburst phase. 
The maximum colour index is attained at the same time as or after the peak brightness, leading to a shallower trajectory during the decay phase. 
This behaviour is observed in the majority of the detected BeT stars in this study and can be explained by the fact that the disk build-up is faster than its dissipation, as pointed out in \citet{Haubois2012}.
Similar loops are also presented in their Figure~15, and it would be possible to use the shape of the loop (e.g. the range of $W1$ and $(W12-W2)$ and the loop width) to constrain the inclination angle of the star because the shape of the model prediction from \citet{Granada2018} for face-on and edge-on disks is different.
A study like this is beyond the scope of this paper, however, and also requires a detailed prediction of the $W1$ and $W2$ magnitude during the disk build-up and dissipation phase.
A combination of the WISE photometry and the observation in the optical could also provide constraints on the inclination angle because the brightness variation in these two bands is different. 

\section{Discussion}
\label{sec:discussion}

The WISE BeT stars detected in this study, 916, 736 of which were newly found, enlarges the number of BeT stars from a few hundred in previous studies to nearly one thousand. 
This large increase in number is achieved by the wide sky-coverage of WISE mission and its long-term photometry in the infrared.
Moreover, $(W1-W2) > \SI{-0.05}{mag}$ can be used as an indicator of the decretion disk.
Previous optical surveys of BeT stars were usually made in one band only.
Whether a star holds a decretion disk when viewed in only one band is hard to determine because a Be star may own a disk before an outburst and still hold a disk after its brightness decreases after the current event.
When WISE epoch photometry is used, it is straightforward to tell whether a star has a disk from its $(W1-W2)$ values because 1) $W1$ and $W2$ data are always available, and 2) the intrinsic colour of early-type stars is $\sim$\SI{-0.05}{mag}. 
This conclusion is also supported by the VDD theory, as we discuss in the next subsection.

\subsection{Comparison with the model prediction}
\label{sec:model-comparison}

Current VDD models can provide the predicted brightness and colour for the decretion disk in different situations, which can be directly compared with the WISE light curves of the BeT stars.
For example, \citet{Granada2018} considered an axisymmetric disk density distribution characterized by two parameters: the density at the disk base ($\rho_0$), and the power-law exponent ($n$) that controls the decrease in density with distance in the equatorial plane from the star.
The authors provided predictions of the decretion disk brightness in 2MASS and WISE bands for several different spectral types, power-law exponents $n$, density $\rho_0$, and inclination angles.
A subset of these predictions, that is, a face-on ($i=18\degree$) and edge-on ($i=84\degree$) case for spectral type B3, is overplotted in Figure~\ref{fig:BeT-CMD}.
The line-connected dots indicate the increase in $\rho_0$ (\num{e-12}--\SI{e-10}{g\,cm^{-3}}) from left to right, and those with parameters that are unlikely to be found are plotted in grey.
Most of the allowed predictions from the model are in the $(W1-W2)$ range of \num{-0.05}--\SI{0.35}{mag}, which is consistent with the $(W1-W2)$ colour range for the BeT stars.
The shape of the prediction curves is slightly different in $W1-W2 > \SI{0.15}{mag}$, which may be used to constrain their $n$ or inclination angle, but this study is beyond the scope of this paper.

The $EW$ of H$\alpha$ was also included in the prediction from \citet{Granada2018}, and it is also plotted in Figure~\ref{fig:BeT-Ha}.
Good consistency is again found between the observation and model, both for the negative correlation and for the fewer points in the area of unlikely parameters.
The scatter in $EW$ can be explained by the different $n$ values of the disks: When the disk is well established (e.g. $W1-W2 > \SI{0.1}{mag}$), a more extended disk (with a larger $n$) reduces the H$\alpha$ emission, while this effect is not obvious when the disk base density is small (i.e. for those with $W1-W2 < \SI{0.1}{mag}$).
Most of the stars also follow the direction of the model lines, indicating that the $\rho_0$ and not the $n$ values change during the Be phenomena.
Future time-series spectroscopic observation with a more complete sampling would provide a more solid conclusion about the variation in the model parameters during this process.

\subsection{Contaminations in the BeT star sample}
\label{sec:contamination}

Although the Be phenomena are defined to occur exclusively in classical Be stars, our BeT sample might include other types of stars.
For example, Herbig Ae/Be stars (HAeBes), which are pre-main-sequence stars that are embedded in circumstellar envelopes, share similar photometry characteristics with classical Be stars. 
They might be included in our extended sample, thereby contributing to the contamination of the final BeT star list. 
The selection criteria for BeT stars require a redder-and-brighter variation ($r < -0.75$), which implies that contaminating Herbig Ae/Be stars must also demonstrate this behaviour.
HAeBes are primarily known for their UXor phenomenon \citep{Waters1998}, in which the varying column density of the circumstellar material between the star and the observer leads to light obstruction, causing the star to appear dimmer.
This in turn results in a redder colour when the star is dimmer, in contrast to the behaviour of the Be phenomena and our selection criteria. 
Hence, the contamination from HAeBe stars is expected to be small.
Only one of the 1361 HAeBes listed in \citet{Vioque2020} is identified as a WISE BeT star, which suggests that HAeBes contaminate only a small fraction of our BeT sample.

We then investigated the BeT stars that were also part of the variability sample presented in \citet{Mei2023}.
As shown in Table~\ref{tab:mei2023_cross}, most of the overlapping stars are CBes.
The contamination of HAeBes (included in type PMS) is again proven to be small.
Most of the burst-type CBes in \citet{Mei2023} are recovered in our BeT sample, indicating that the variation of the stars in this type is caused by the Be phenomenon.
The lower recovery rate for other types of CBes mainly arises because their magnitude and colour are not negatively correlated, that is, some of them are the group 2 stars classified in section~\ref{sec:selecting-BeT}.
This difference implies that limiting the $r$ value is important to distinguish the Be phenomena light curve from other variations.
We also note that the selection of the epoch photometry is stricter in this study, and some of the variable CBes in \citet{Mei2023} may therefore be excluded due to insufficient binned photometry.

Table~\ref{tab:bep-catalogue} lists the main object type of BeT stars from the Simbad database.
Most of the 769 stars with an object type are labelled Be stars (262), emission line stars (277), variable stars (43), or stars (157).
The remaining 30 stars mainly consist of young stellar objects or candidates. 
Two Cepheids are included in the BeT sample.
The behaviour of the first Cepheid, 06355760+0114399, is similar to that of the Be phenomena, with an increase in brightness from March 2010 to March 2016, and a decay until March 2022.
It is classified as a Cepheid based on the optical light curve observed by the All-Sky Automated Survey for Supernovae survey (ASAS-SN; \citealt{Kochanek2017}).
However, the ASAS-SN optical light curve is similar to its WISE light curve without any periodic characteristic, and we therefore conclude that it is a Be star and not a Chepheid.
The second Cepheid (18231006+0130191 from \citet{Zhang-YJ2022}; plotted as dotted grey lines in Figure~\ref{fig:BeT-CMD}), however, shows a semi-periodical light curve that does not resemble the Be phenomenon.
Its ASAS-SN light curve has a period of \SI{20.99}{days}, similar to the length of each visit ($\sim$\SI{15}{days}) for the WISE survey. 
The visit binned photometry in this case cannot trace the Cepheid light curve correctly. 
We therefore removed this star from our BeT sample.
We recommend checking their optical light curve for the BeT stars having some suspicious light curves in WISE bands.

\begin{table*}
    \caption{Number of BeT stars in common with \citet{Mei2023}, with the numbers of CBe stars in their study given in parentheses.}
    \centering
    \begin{tabular}{lcccccccc}
    \hline\hline
    \multirow{2}{*}{Sample} & \multicolumn{3}{c}{Secular} & & \multicolumn{3}{c}{Stochastic} & Total \\
    \cline{2-4} \cline{6-8}
    & sin & sin+linear & linear & & burst & drop & irregular \\
    PMS & 1 & 0 & 0 & & 1 & 0 & 0 & 2\\
    Uncertain & 1 & 1 & 0  & & 0 & 0 & 3 & 5\\
    CBe & 23 (94) & 19 (92) & 19 (91) & & 11 (18) & 8 (36) & 26 (128) & 106 (459)\\
    \hline
         &  \\
    \end{tabular}
    \label{tab:mei2023_cross}
\end{table*}

We expect that the contamination from other types of variable stars is marginal.
One hundred and seventeen BeT stars have a confidence classification (\texttt{best\_class\_score} $> 0.5$) in the catalogue of \citet{Rimoldini2023}, who classified 12.4 million variable sources into 25 classes using Gaia epoch photometry.
Seventy percent of them are classified as Be stars, and an additional 20\% are rotational variables ($\alpha^2$ Canum Venaticorum variables or RS Canum Venaticorum variables).
The remaining 10\% of the stars mainly consist of $\delta$ Scuti variables, which also have similar spectral types as BeT stars.
Consequently, the contamination of other variables, including symbiotic stars or cataclysmic variables, is negligible. 

\subsection{BeT stars in open clusters}

BeT stars are expected to be found in open clusters because these clusters host many early-type stars.
As stated in section~\ref{sec:intro}, \citet{Granada2018} found that about half of the Be stars in young open clusters are active or have $W1-W2$ colours redder than \SI{0.05}{mag}.
The number of BeT stars in clusters would be lower, however.
A star with a decretion disk may stay in this phase for a long time and therefore show no brightness or colour variation.
Our detection criteria therefore do not classify these stars as BeT stars.
We matched the positions of BeT stars with recent open cluster catalogues, that is, \citet{Cantat-Gaudin2020} and \citet{Pang2022}.
Thirty-five BeT stars are found in 26 open clusters, as listed in Table~\ref{tab:BeT_OC}.
The number of BeT stars in each cluster is small, with a maximum of 5 for NGC\,663 and a minimum of 1 for 27 clusters.
Slightly more BeT stars lie in relatively young clusters (age $< 45$\,Myr) than in older ones.
The main-sequence turn-off point for a single-population star with an age of \SI{45}{Myr} is about \SI{18000}{K}, which is close to the peak $\Teff$ distribution of our BeT stars.
When we assume that the $\Teff$ distribution of cluster BeT stars is the same as for the field stars, the larger number of BeT stars in younger cluster can simply be explained as the result of stellar evolution. Most of the OB stars evolve away from the main sequence and are therefore beyond our detection range.

\begin{table*}
    \caption{Open clusters hosting WISE BeT stars, with their age from \citet{Cantat-Gaudin2020} and the number of BeT stars.}
    \centering
    \begin{tabular}{ccc|ccc}
    \hline\hline
    Cluster & Age & \# of BeT stars & Cluster & Age & \# of BeT stars \\
     & (Myr) & & & (Myr) & \\
    \hline
    Berkeley 87 &        8.3 &       1 &      NGC 7510 &       19.5 &       1 \\
    Alessi 43 &       11.5 &       1 &       NGC 457 &       20.9 &       2 \\
     NGC 4755 &       12.0 &       1 &      FSR 0904 &       21.9 &       1 \\
     NGC 2244 &       12.6 &       2 &      NGC 3766 &       22.9 &       1 \\
      Stock 8 &       14.5 &       1 &       NGC 663 &       29.5 &       5 \\
      NGC 869 &       15.1 &       2 &      Pismis 1 &       29.5 &       1 \\
      NGC 659 &       15.5 &       1 & Collinder 272 &       45.7 &       1 \\
      IC 1848 &       15.8 &       1 & Collinder 258 &       97.7 &       1 \\
    Gulliver 6 &       16.6 &       1 &      Stock 23 &      109.6 &       1 \\
      King 10 &       17.4 &       1 &      NGC 7790 &      128.8 &       1 \\
      UBC 320 &       17.8 &       1 &      NGC 5662 &      199.5 &       1 \\
      UBC 191 &       18.2 &       1 &      NGC 1582 &      234.4 &       1 \\
    Collinder 107 &       18.2 &       1 &      NGC 1857 &      251.2 &       1 \\
    \hline
    \end{tabular}
    \label{tab:BeT_OC}
\end{table*}

\section{Conclusion}
\label{sec:conclusion}

Outbursts caused by Be phenomena were detected in 916 early-type stars in the mid-infrared bands. 
These BeT stars exhibit an increased brightness in the $W1$ and $W2$ bands and reddening in $(W1-W2)$, or vice versa. 
BeT star light curves can be identified using criteria based on $W1$ magnitude and $(W1-W2)$, and $W1-W2 = \SI{-0.05}{mag}$ is a criterion for the presence of a decretion disk. 
The connection between brightness variation and Be phenomena is confirmed: 1) $(W1-W2)$ values during outbursts are typically below $\sim \SI{0.35}{mag}$, consistent with model predictions, and 2) the emission of the H$\alpha$ line (indicating circumstellar disks) correlates with $(W1-W2)$. 
The median outburst and quiescent phase duration are about 500 days (those with fewer than 180\,days were not sampled due to the large cadence of the light curve), and behaviour differences in these two phases can be seen as a clockwise loop in epoch-CMD.
About 25 open clusters host 33 of the BeT stars we found, with a large portion in young clusters (age $< \SI{45}{Myr}$).
The current data set and future data from the WISE survey will provide a better understanding of the physical processes involved in the Be phenomenon.


\begin{acknowledgements}
MJ thanks Xiaodian Chen and Xiaoying Pang for the useful discussion.
BJ thanks the support from the National Key R\&D Program of China No. 2019YFA0405503.
This publication makes use of data products from the Wide-field Infrared Survey Explorer, which is a joint project of the University of California, Los Angeles, and the Jet Propulsion Laboratory/California Institute of Technology, funded by the National Aeronautics and Space Administration.
This publication also makes use of data products from NEOWISE, which is a project of the Jet Propulsion Laboratory/California Institute of Technology, funded by the Planetary Science Division of the National Aeronautics and Space Administration.
The light curve extraction was enabled by resources provided by the Swedish National Infrastructure for Computing (SNIC) at the PDC Center for High Performance Computing, KTH Royal Institute of Technology, partially funded by the Swedish Research Council through grant agreement no. 2018-05973.
This research has made use of the Simbad database, operated at CDS, Strasbourg, France.
\end{acknowledgements}

%
  \bibliographystyle{aa} 
  \bibliography{refs} 
%






   
  



\end{document}